\documentclass{nature_plusfigure}
\usepackage{graphicx}
\usepackage{amsmath,xcolor,url}
\usepackage{multirow}
\usepackage{longtable}
\usepackage{amssymb}
\usepackage{hyperref}
\graphicspath{{./}{figures/}}
\usepackage{lineno} 
\def\apj{Astrophys.\ J.}
\def\pasp{Publ. Astron. Soc. Pac}
\def\mnras{	Mon. Not. R. Astron. Soc.}
\def\prd{Phys. Rev. D}
\def\aap{Astron. Astrophys}
\def\pasa{Publ. Astron. Soc. Aust.}
\def\apjl{Astrophys.\ J. Lett.}
\def\araa{Annu. Rev. Astron. Astrophys.}
\def\prl{Phys. Rev. Lett.}
\def\aj{Astron. J}
\def\apjs{Astrophys.\ J. Suppl.}
\def\ssr{Space Sci. Rev}

\newcommand{\arcdeg}{\mbox{$^\circ$}}
\newcommand{\arcsec}{$^{\prime\prime}$}

\newcommand{\Msol}{\mbox{$M_{\odot}$}}

\newcounter{firstbib} % for the ref split between main text and SI

\title{A tidal disruption event coincident with a high-energy neutrino}

\author{Robert Stein$^{*1, 2}$,
	Sjoert van Velzen$^{*3, 4, 5}$, 
	Marek Kowalski$^{*1, 2, 6}$,
	Anna Franckowiak$^{1, 2}$,
	Suvi Gezari$^{4, 7}$,
	James C. A. Miller-Jones$^{8}$,
	Sara Frederick$^{4}$,
	Itai Sfaradi$^{9}$,
	Michael F. Bietenholz$^{10, 11}$,
	Assaf Horesh$^{9}$,
	Rob Fender$^{12,13}$,
	Simone Garrappa$^{1,2}$,
	Tom\'as Ahumada$^{4}$,
	Igor Andreoni$^{14}$,
	Justin Belicki$^{15}$,
	Eric C. Bellm$^{16}$,
	Markus B\"ottcher$^{17}$,
	Valery Brinnel$^{2}$,
	Rick Burruss$^{15}$,
	S. Bradley Cenko$^{7,18}$,
	Michael W. Coughlin$^{19}$,
	Virginia Cunningham$^{4}$,
	Andrew Drake$^{14}$,
	Glennys R. Farrar$^{3}$,
	Michael Feeney$^{15}$,
	Ryan J. Foley$^{20}$,
	Avishay Gal-Yam$^{21}$,
	V. Zach Golkhou$^{16,22}$,
	Ariel Goobar$^{23}$,
	Matthew J. Graham$^{14}$,
	Erica Hammerstein$^{4}$,
	George Helou$^{24}$,
	Tiara Hung$^{20}$,
	Mansi M. Kasliwal$^{14}$,
	Charles~D.~Kilpatrick$^{20}$,
	Albert K. H. Kong$^{25}$,
	Thomas Kupfer$^{26}$,
	Russ R. Laher$^{24}$,
	Ashish A. Mahabal$^{14,27}$,
	Frank J. Masci$^{24}$,
	Jannis Necker$^{1,2}$,
	Jakob Nordin$^{2}$,
	Daniel A. Perley$^{28}$,
	Mickael Rigault$^{29}$,
	Simeon Reusch$^{1,2}$,
	Hector Rodriguez$^{15}$,
	C\'{e}sar~Rojas-Bravo$^{20}$,
	Ben Rusholme$^{24}$,
	David L. Shupe$^{24}$,
	Leo P. Singer$^{18}$,
	Jesper Sollerman$^{30}$,
	Maayane T. Soumagnac$^{21,31}$,
	Daniel Stern$^{32}$,
	Kirsty Taggart$^{28}$,
	Jakob van Santen$^{1}$,
	Charlotte Ward$^{4}$,
	Patrick Woudt$^{13}$,
	Yuhan Yao$^{14}$
}

\begin{document}

\maketitle
\newline

\begin{affiliations}
 \item {Deutsches Elektronen Synchrotron DESY, Platanenallee 6, 15738 Zeuthen, Germany}
 \item {Institut f{\"u}r Physik, Humboldt-Universit{\"a}t zu Berlin, D-12489 Berlin, Germany}
\item Center for Cosmology and Particle Physics, New York University, NY 10003, USA
\item Department of Astronomy, University of Maryland, College Park, MD 20742, USA
\item Leiden Observatory, Leiden University, PO Box 9513, 2300 RA Leiden, The Netherlands
\item {Columbia Astrophysics Laboratory, Columbia University in the City of New York, 550 W 120th St., New York, NY 10027, USA}
 \item Joint Space-Science Institute, University of Maryland, College Park, MD 20742, USA
 \item International Centre for Radio Astronomy Research -- Curtin University, Perth, Western Australia, Australia
 % Itai Sfaradi and Assaf Horesh
 \item {Racah Institute of Physics, The Hebrew University of Jerusalem, Jerusalem 91904, Israel} % 
 % M. Bietenholz, A. de Witt
 \item{Hartebeesthoek Radio Astronomy Observatory, SARAO, PO Box 443, Krugersdorp, 1740, South Africa }
 \item{Department of Physics and Astronomy, York University, Toronto, M3J~1P3, Ontario, Canada}
% Rob Fender, P. Wouldt
\item Astrophysics, Department of Physics, University of Oxford, Keble Road, OX1 3RH, UK
\item Department of Astronomy, University of Cape Town, Private Bag X3, Rondebosch, 7701, South
Africa
%  I. Andreoni, Y. Yao, M. Graham
\item {Division of Physics, Mathematics, and Astronomy, California Institute of Technology, 1200 East California Blvd, MC 249-17, Pasadena, CA 91125, USA}
% J. Belecki, R. Burrus, M. Feeney, H. Rodriguez
\item{Caltech Optical Observatories, California Institute of Technology, Pasadena, CA  91125, USA}
% E. Bellm + V. Zach Golkhou
\item{DIRAC Institute, Department of Astronomy, University of Washington, 3910 15th Avenue NE, Seattle, WA 98195, USA}
% M. Boettcher
\item{Centre for Space Research, North-West University, Potchefstroom, 2531,
South Africa}
% S. B. Cenko, L. P. Singer, 
\item{Astroparticle Physics Laboratory, NASA Goddard Space Flight Center, Mail Code 661, Greenbelt, MD 20771, USA}
%  M. Coughlin:
\item{School of Physics and Astronomy, University of Minnesota,
Minneapolis, Minnesota 55455, USA}
% Ryan Foley, Tiara Hung, 
\item{Department of Astronomy and Astrophysics, University of California, Santa Cruz, California, 95064, USA}
% Gal-Yam, Soumagnac
\item{Department of Particle Physics and Astrophysics, Weizmann Institute of Science, 234 Herzl St, Rehovot 76100, Israel}
% V. Zach Golkhou
\item{The eScience Institute, University of Washington, Seattle, WA 98195, USA}
% A. Goobar
\item{The Oskar Klein Centre, Department of Physics,  Stockholm University,
AlbaNova, SE-106 91 Stockholm, Sweden}
% Helou, Laher, Masci, Rusholme, Shupe
\item{IPAC, California Institute of Technology, 1200 E. California Blvd, Pasadena, CA 91125, USA}
% A. K. H. Kong
\item{Institute of Astronomy, National Tsing Hua University, No. 101 Section 2 Kuang-Fu Road, Hsinchu 30013, Taiwan}
% T. Kupfer
\item{Kavli Institute for Theoretical Physics, University of California, Santa Barbara, CA 93106, USA}
% A. Hamabal (+ caltech)
\item{Center for Data Driven Discovery, California Institute of Technology, Pasadena, CA 91125, USA}
% D. Perley + K. Taggart
\item{Astrophysics Research Institute, Liverpool John Moores University, 146 Brownlow Hill, Liverpool L3 5RF, UK
}
% Mickael Rigault
\item{Universit\'e Clermont Auvergne, CNRS/IN2P3, Laboratoire de Physique de Clermont, F-63000 Clermont-Ferrand, France}
% J. Sollerman
\item{The Oskar Klein Centre, Department of Astronomy,  Stockholm University,
AlbaNova, SE-106 91 Stockholm, Sweden}
% M. Soumagnac
\item{Lawrence Berkeley National Laboratory, 1 Cyclotron Road, Berkeley, CA 94720, USA}
% D. Stern
\item{Jet Propulsion Laboratory, California Institute of Technology, 4800 Oak Grove Drive, Pasadena, CA 91109, USA}
\end{affiliations}

\begin{abstract}
\textbf{Cosmic neutrinos provide a unique window into the otherwise-hidden mechanism of particle acceleration in astrophysical objects. A flux of high-energy neutrinos was discovered in 2013, and the IceCube Collaboration recently reported the likely association of one high-energy neutrino with a flare from the relativistic jet of an active galaxy pointed towards the Earth. However a combined analysis of many similar active galaxies revealed no excess from the broader population, leaving the vast majority of the cosmic neutrino flux unexplained. Here we present the likely association of a radio-emitting tidal disruption event, AT2019dsg, with a second high-energy neutrino. AT2019dsg was identified as part of our systematic search for optical counterparts to high-energy neutrinos with the Zwicky Transient Facility. The probability of finding any coincident radio-emitting tidal disruption event by chance is 0.5\%, while the probability of finding one as bright in bolometric energy flux as AT2019dsg is 0.2\%. Our electromagnetic observations can be explained through a multi-zone model, with radio analysis revealing a central engine, embedded in a UV photosphere, that powers an extended synchrotron-emitting outflow.
This provides an ideal site for PeV neutrino production. Assuming that the association is genuine, our observations suggest that tidal disruption events with mildly-relativistic outflows contribute to the cosmic neutrino flux.}
\end{abstract}

On 2019 October 1, the IceCube Neutrino Observatory\cite{icecube_detector} reported the detection of a $\sim$0.2 PeV neutrino, IC191001A, with an estimated 59\% probability of being of astrophysical origin based solely on reconstructed energy\cite{stein:gcn25913}. Seven hours later, the direction of the incoming neutrino was observed by the Zwicky Transient Facility (ZTF)\cite{2019PASP..131a8002B} as part of our neutrino follow-up program. The data were processed by our multi-messenger pipeline (see Methods), which performs searches for extragalactic transients in spatial and temporal coincidence with high-energy neutrinos\cite{2007APh....27..533K}, and the radio-emitting tidal disruption event AT2019dsg was identified as a candidate neutrino source.

Tidal Disruption Events (TDEs) are rare transients that occur when stars pass close to supermassive black holes (SMBHs). Studies have suggested that TDEs are sources of high-energy neutrinos and ultra-high energy cosmic rays\cite{2009ApJ...693..329F,2017MNRAS.469.1354D, 2019ApJ...886..114H}, in particular the subset of TDEs with relativistic particle jets\cite{2014arXiv1411.0704F,2017ApJ...838....3S,2016PhRvD..93h3005W,2017PhRvD..95l3001L}. Those TDEs with non-thermal emission are considered the most likely to be sources of high-energy neutrinos. AT2019dsg was thus quickly identified as a promising candidate neutrino source\cite{2019ATel13160....1S}. Given that there are typically $\lesssim$2 radio-emitting TDEs in the entire northern sky at any one time, we find that in the 80 sq. deg. of sky observed during the eight neutrino follow-up campaigns by ZTF up to March 2020, the probability of finding a radio-detected TDE-neutrino association by chance is $<$0.5\%. With the second highest bolometric energy flux of all seventeen TDEs detected by ZTF, the probability of finding a TDE at least as bright as AT2019dsg by chance is just 0.2\%. These calculations are valid for any isotropic distribution, and therefore quantify the probability that the AT2019dsg-IC191001A association would arise from atmospheric backgrounds. Our program targets four neutrino population hypotheses\cite{2019PASP..131g8001G}, of which the greatest sensitivity is for TDEs (see Methods). Thus, although not directly reflected in the calculation, the impact of multiple hypothesis tests on these estimates would be modest. While an atmospheric origin  for the IC191001A-AT2019dsg association cannot be excluded, the improbability of chance temporal and spatial coincidence substantially reinforces the independent energy-based evidence of an astrophysical origin for IC191001A, and indicates that any atmospheric origin is unlikely.

\begin{figure}
\includegraphics[width=\textwidth]{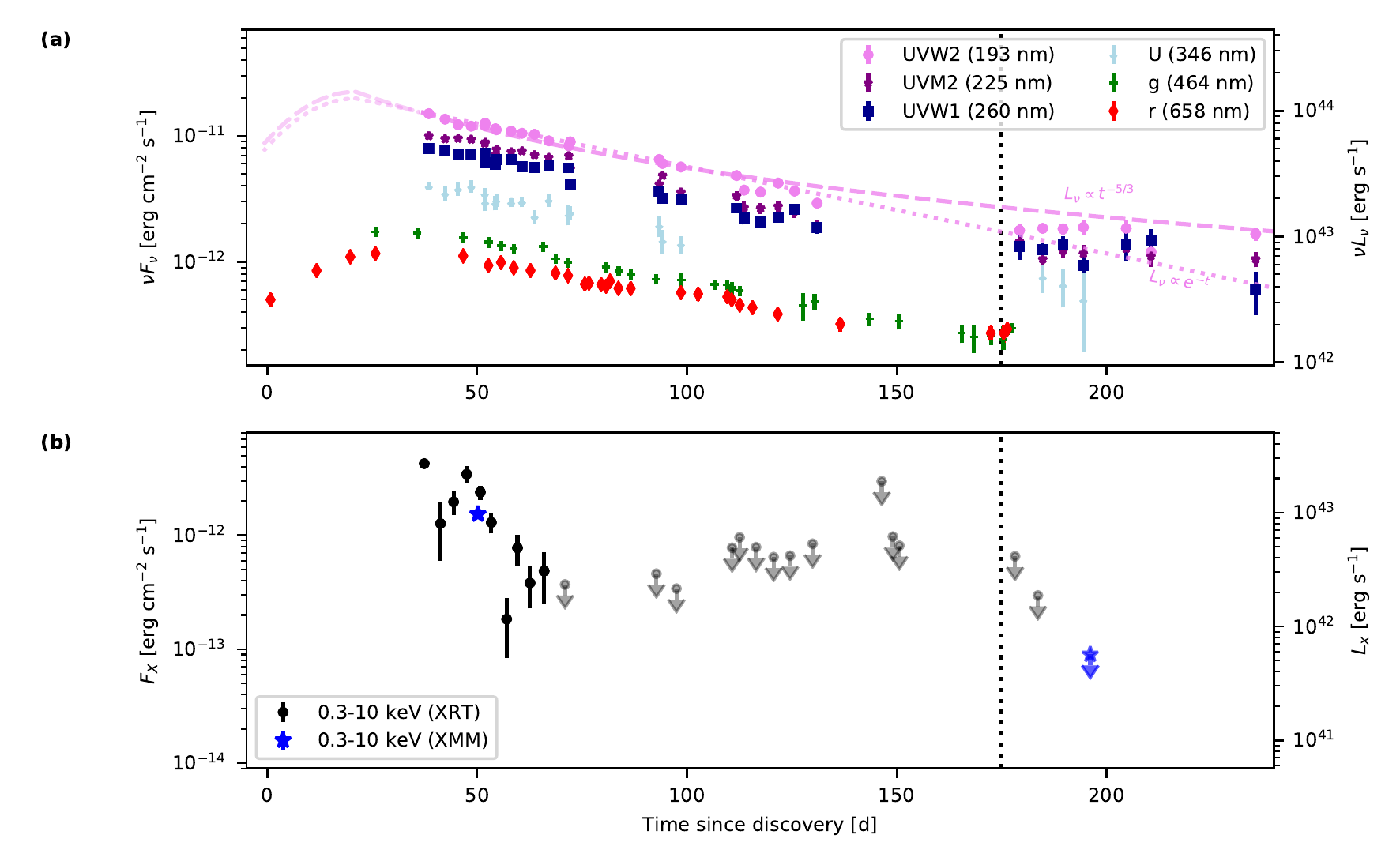}
\caption{\textbf{Multi-wavelength lightcurve of AT2019dsg.} Error bars represent 1$\sigma$ intervals. The upper panel, a, shows the optical photometry from ZTF (in green and red), alongside UV observations from \textit{Swift}-UVOT (in blue, purple and pink). 
%The apparent late-time plateau  is a factor of 10 brighter in UVW2 than the pre-disruption baseline of the host galaxy. 
The late-time UV observations show an apparent plateau which is not captured by a single power-law decay. The dashed pink line illustrates a canonical t$^{-5/3}$ power law, while the dotted pink line illustrates an exponentially-decaying lightcurve. Neither model describes the UV data well.
The lower panel, b, shows the integrated X-ray energy flux, from observations with \textit{Swift}-XRT (in black) and \textit{XMM-Newton} (in blue), in the energy range 0.3-10 keV. Arrows indicated 3$\sigma$ upper limits.  The vertical dotted line illustrates the arrival of IC191001A.}
\label{fig:lightcurve}
\end{figure}

AT2019dsg was discovered\cite{2019TNSTR.615....1N} by ZTF on 2019 April 9, and was classified as a TDE on the basis of its optical spectrum\cite{2019ATel12752....1N}. This spectrum showed a redshift of $z=0.051$, implying a luminosity distance $D_{\rm L} \approx$ 230 Mpc assuming a flat cosmology with $\Omega_{\Lambda}$ = 0.7 and $H_{0}$ = 70 km s$^{-1}$ Mpc$^{-1}$. The optical/UV continuum of AT2019dsg is well described by a single blackbody photosphere with a near-constant temperature\cite{2020arXiv200101409V} of $10^{4.59 \pm 0.02}$~K and radius of $10^{14.59 \pm 0.03}$~cm. The peak luminosity of $10^{44.54 \pm 0.08}$ erg\,s$^{-1}$ is in the top 10\% of the 40 known optical TDEs to date\cite{2020arXiv200101409V}, and the temperature is in the top 5\%. The late-time evolution is consistent with the rapid formation of an accretion disk\cite{2019ApJ...878...82V,2020MNRAS.492.5655M} (see Figure \ref{fig:lightcurve}), which would be expected on these relatively short timescales for disruptions around higher-mass SMBHs. Indeed the total mass of the host galaxy of AT2019dsg is in the top 10$\%$ of all optical TDE hosts. Assuming 50$\%$ of the host mass is in the bulge, we estimate\cite{2013ApJ...764..184M} a black hole mass of $\sim 3 \times 10^{7} \Msol$.

% By the time of the neutrino detection, the optical/UV luminosity appeared to have reached a plateau (see Figure \ref{fig:lightcurve}). Such plateaus are common in TDEs and interpreted as emission from the outer part of an accretion disk\cite{2019ApJ...878...82V,2020MNRAS.492.5655M}, but typically occur a few years after peak. The rapid appearance of an accretion-disk plateau would be expected for disruptions around higher-mass SMBHs. Indeed the total mass of the host galaxy of AT2019dsg is in the top 10$\%$ of all optical TDE hosts. Assuming 50$\%$ of the host mass is in the bulge, we estimate\cite{2013ApJ...764..184M} a black
% hole mass of $\sim 3 \times 10^{7} \Msol$.

AT2019dsg was also detected in X-rays, beginning 37 days after discovery. Though the first X-ray observation indicated a bright source, with a high X-ray to optical ratio of $L_{\rm X}/L_{\rm opt} \sim $0.1, this X-ray flux faded extremely rapidly, as shown in Figure \ref{fig:lightcurve}. This rate of decline is unprecedented, with at least a factor of 50 decrease in X-ray flux over a period of 159 days. Similar to the optical/UV emission, the observed X-ray spectrum is consistent with thermal emission, but from a blackbody of temperature $10^{5.9}$ K ($0.072 \pm 0.005$ keV) and, assuming emission from a circular disk, a radius $\sim 2 \times 10^{11}$ cm. As for most X-ray-detected TDEs\cite{2017ApJ...838..149A, 2019MNRAS.487.4136W,2019ApJ...872..198V}, the blackbody radius appears much smaller than the Schwarzschild radius ($R_{\rm S} \sim 10^{13}$~cm) inferred from the galaxy scaling relation\cite{2013ApJ...764..184M}.  X-ray emission is generally expected to arise close to the Schwarzschild radius. Small emitting areas can arise from an edge-on orientation, because the relativistic velocities at the inner disk can Doppler boost a large area of the disk out of the X-ray band.  Since our observations probe close to the Wien tail of the spectrum, a small temperature decrease due to absorption would also yield a significantly underestimated blackbody radius and luminosity\cite{2019ApJ...872..198V}. The exponential decrease of the flux could be caused by cooling of the newly-formed TDE accretion disk\cite{2020MNRAS.492.5655M} or increasing X-ray obscuration.

\begin{figure}
\includegraphics[width=\textwidth]{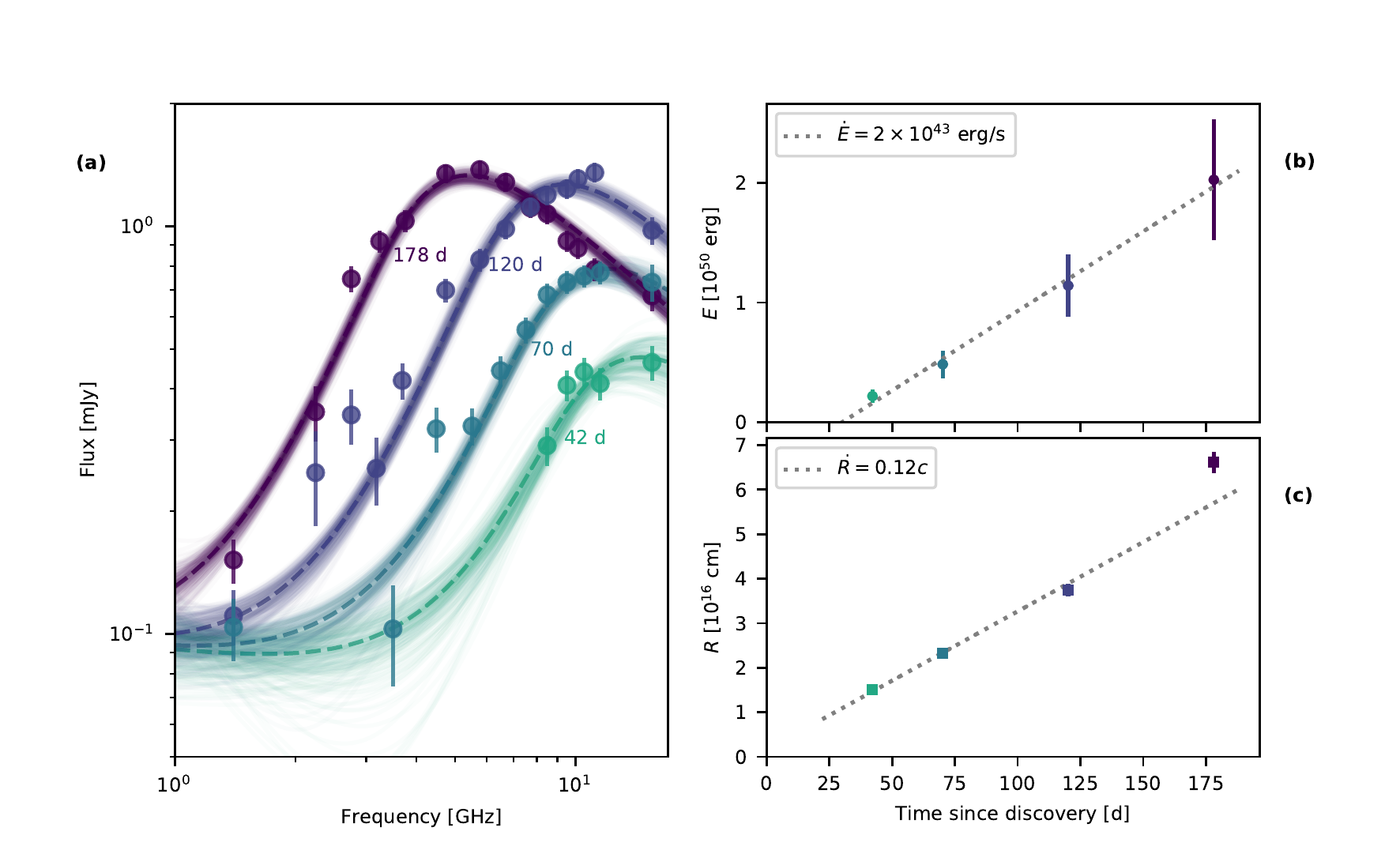}
\caption{\textbf{Synchrotron analysis of AT2019dsg.} Panel (a) shows radio measurements from MeerKAT (1.3 GHz), VLA (2-12 GHz), and AMI (15.5 GHz), at four epochs with times listed relative to the first optical detection. The coloured lines show samples from the posterior distribution of synchrotron spectra fitted to the measurements at each epoch, and the dashed lines trace the best-fit parameters for that epoch. The free parameters are the electron power-law index ($p=2.9\pm 0.1$), the host baseline flux density, plus the magnetic field and radius for each epoch. Panel (b) shows the energy at each epoch for a conical outflow geometry with an half-opening angle of 30\arcdeg. The dotted line indicates a linear increase of energy.  Panel (c) shows the corresponding radius for each epoch, with a dotted line illustrating a linear increase. Error bars represent 1$\sigma$ intervals.}
\label{fig:radio_spectrum}
\end{figure}

Radio observations shown in Figure \ref{fig:radio_spectrum} reveal a third distinct spectral component, namely synchrotron emission from non-thermal electrons. We model this emission with a  conical geometry as expected for outflows (e.g., jets or winds) that are launched from---and collimated by---the inner parts of flared accretion disks that emit close to the Eddington limit.
Given that electrons are typically accelerated with much lower efficiency than protons in astrophysical accelerators\cite{2012A&A...538A..81M}, we assume that they carry 10\% of the energy carried by relativistic protons ($\epsilon_{e} = 0.1$). We further assume that the magnetic fields carry 0.1\% of the total energy ($\epsilon_{B} = 10^{-3}$), as indicated by radio observations of other TDEs\cite{2018ApJ...854...86E} and supernovae\cite{2013MNRAS.436.1258H}. We note that the opening angle for the outflow is largely unconstrained. For a half-opening angle, $\phi$, of 30\arcdeg\ we find $R = 1.5 \times 10^{16}$ cm in our first epoch (41 days after discovery), increasing to $R = 7 \times 10^{16}$~cm shortly after the neutrino detection (177 days after discovery). These radii scale\cite{2013ApJ...772...78B} as $R \propto [1-\cos(\phi$)]$^{-8/19}$. The implied expansion velocity is roughly constant at $v/c = \dot{R}/c = 0.12 \pm 0.01 $ during the first three epochs, with a significant ($>3\sigma$) acceleration to $v/c = 0.21 \pm0.02$ for the last epoch. These are the velocities of the synchrotron-emitting region, and thus provide a lower limit to the velocity at the base of the outflow.  Indeed even the hotspots of relativistic jets from active galaxies that are frustrated by gas in their host galaxy are typically observed\cite{2003PASA...20...69P} to have subrelativistic expansion velocities of $\sim0.1c$. 

The inferred outflow energy, $E$, shows a linear increase from $2.5 \times 10^{49}$~erg to $2 \times 10^{50}$~erg (Figure \ref{fig:radio_spectrum}), which would not be expected from models of TDE radio emission which involve a single injection of energy\cite{2016ApJ...819L..25A,2016ApJ...827..127K}. The constant increase of energy implies a constant injection rate at the base of the outflow of approximately $2\times10^{43}$\,erg\,s$^{-1}$. While some scenarios can yield an increase in inferred energy from a single energy injection, none of these are consistent with the full set of observed properties. First, a single ejection with a range of velocities could explain the observed linear increase of energy with time (the slower ejecta arrive later), but is incompatible with the increasing velocity. Second, an increase of the efficiency for conversion of Poynting luminosity to relativistic particles is unlikely because the target density that is available to establish this conversion is decreasing.  And finally, an apparent increase of the inferred energy due to an increase of solid angle that emits to our line of sight is only expected for relativistic outflows that decelerate. Instead, for AT2019dsg, the observations suggest the presence of a central engine that yields continuous energy injection through a coupling of accretion power to the radio emission\cite{2018ApJ...856....1P}, with acceleration in the final radio epoch due to a decrease in the slope of the ambient matter density profile. 

\subsection*{Neutrino emission from AT2019dsg.} With this strong evidence for three distinct emission zones derived purely from multi-wavelength observations, we consider whether this picture is consistent with AT2019dsg being the source of the neutrino IC191001A. In particular, neutrino production requires protons to be accelerated to sufficiently high energies, and to collide with a suitably abundant target. The detection of a single high-energy neutrino implies a mean expectation in the range $0.05 < N_{\nu, \textup{tot}} < 4.74$ at 90\% confidence, where $N_{\nu, \textup{tot}}$ is the cumulative neutrino expectation for all TDEs that ZTF has observed, while for an individual object the expectation will be significantly lower\cite{2019A&A...622L...9S}. AT2019dsg emits $f_{\rm bol} \sim 0.16$ of the population bolometric energy flux, and if we take this as a proxy for neutrino emission, we would expect $0.008 \lesssim N_{\nu} \lesssim 0.76$ for this source.

% Accounting for the large Eddington bias that arises from the Poisson-dominated emission of single high-energy neutrinos\cite{2019A&A...622L...9S}, the TDE should reasonably have sufficient flux to produce an expected number of high-energy neutrino alerts $N_{\nu} \sim$ 0.01-1. 

Radio observations confirm that particle acceleration is indeed occurring, and that this continues without decline through to the detection of the neutrino at $\sim$180 days post-discovery. Given that neutrinos typically take a fraction $\eta_{p\nu} \sim 0.05$ of the parent proton energy, our accelerator must be capable of accelerating protons to at least 4 PeV. We evaluate the Hillas criterion\cite{1984ARA&A..22..425H} that the proton Larmor radius be less than the system size, to determine whether this is possible. We use  our estimates for conditions in the synchrotron zone at the time of neutrino detection, with $B \sim$ 0.07 G and $R \sim 7 \times 10^{16}$ cm for the near-contemporaneous radio epoch. Taking this as a baseline, we find a maximum proton energy of $\sim$160 PeV, far in excess of our requirements. The Hillas criterion can also be satisfied within the engine that powers the radio-emitting outflow because the product $BR$ is not expected to decrease at smaller radii (e.g. $B \propto R^{-1}$ for a toroidal configuration). 

The target for neutrino production can be either photons (p$\gamma$ interactions) or protons (pp interactions). For a photon target, neutrino production occurs above an energy determined by the mass of the $\Delta$ resonance, $m_{\Delta}$. For a thermal spectrum, of temperature $T$, we then find $\epsilon_{\nu} \sim \eta_{p\nu}[(m_{\Delta}^{2} - m_{p}^{2})/ 4 \epsilon_\gamma ] \approx 0.3 \times \left(T/10^{5} \rm K \right)^{-1} \rm PeV$. For the UV photosphere of the TDE, we find $\epsilon_\nu \sim$ 0.8 PeV, while for the compact X-ray source, we find $\epsilon_\nu \sim$ 0.05 PeV. Both of these values are compatible with the observed neutrino, for which there is a typical uncertainty of one energy decade\cite{2018Sci...361.1378I}, so either photon field could serve as a target. For the UV photosphere, we find that the mean free path for the parent proton of a PeV neutrino ($\sim2 \times 10^{13}$ cm, see Supplementary Information; SI) is much smaller than the photosphere radius, so the UV photosphere is indeed optically thick. At smaller radii, the X-rays would overtake the UV photons as dominant scattering targets. 

\begin{figure}
\includegraphics[width=\textwidth]{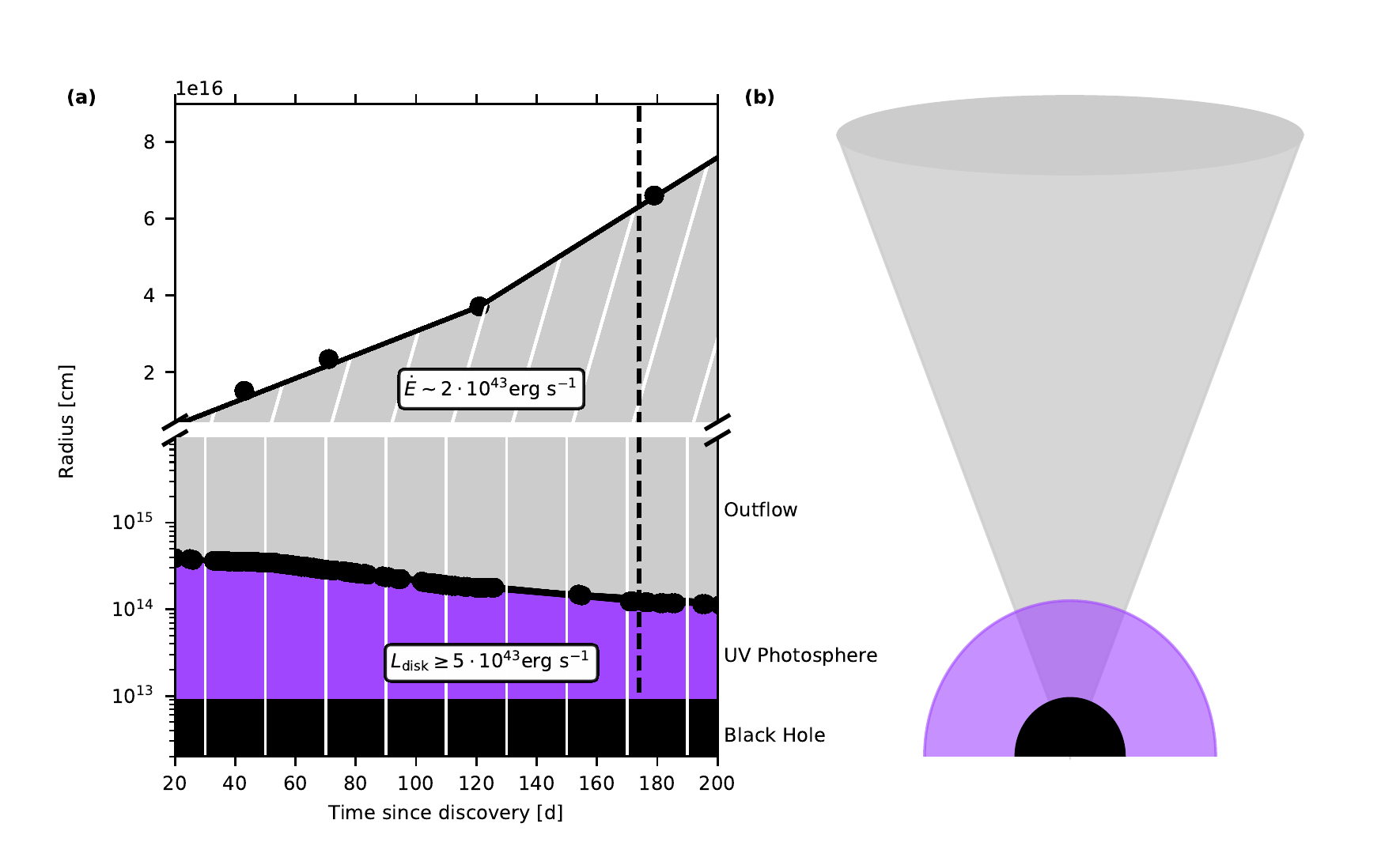}
\caption{\textbf{Diagram of the three emission zones in AT2019dsg.} Panel (a) shows the temporal evolution of the three emission regions, while the geometry of these same regions is illustrated in panel (b). The size of the region responsible for radio emission (in gray), as well as the blackbody radius for the UV-emission (in purple) is derived from data. The Schwarzschild radius is plotted in black for a black hole mass of $3 \times 10^7 \Msol$. The white lines in panel (a) represent a continuous outflow with velocity $c$. 
}
\label{fig:diagram}
\end{figure}

In the multi-zone model, shown in Figure \ref{fig:diagram}, the thermal photons provide a guaranteed target for pion production. However hadrons could in principle also serve as a target, leading us to consider a single-zone scenario in which the protons are accelerated at the same location as the synchrotron-emitting electrons, with the neutrino spectrum following the same intrinsic energy power law as the protons and electrons. For pp neutrino production, high target densities of $n_{\rm p}\sim 1/(\sigma_{\rm pp} R)\sim 10^8$~cm$^{-3}$ would be required for efficient production of neutrinos, where $\sigma_{\rm pp}$ is the proton-proton cross section and $R \sim 10^{17}$~cm is the characteristic size of the radio region at the time of neutrino production. This high density could be provided by the unbound stellar debris, although this component moves with a typical maximum velocity\cite{2016ApJ...827..127K} of $0.05\,c$ , and therefore the majority of this debris would have to be swept up with the outflow. Alternatively, the density could be provided by pre-existing gas, although since this gas orbits in the sphere of influence of the black hole, it would be challenging to satisfy the upper limits on pre-disruption accretion. 

To obtain the expected neutrino flux from this source we have to estimate the energy carried by protons ($E_{\textup{p}}$) that are accelerated above the energy threshold needed to produce high-energy neutrinos. 
The outflow energy of $2 \times 10^{50}$~erg that we derived from the radio observations (Figure~\ref{fig:radio_spectrum}) represent a lower bound to the energy that is available for particle acceleration in a central engine.  Indeed, the total energy budget for a TDE is set by the mass of the disrupted star, with $E_{\textup{TDE}} \sim (1/2)\, 0.1 \, \Msol \, c^{2} \sim 10^{53}$ erg for a solar-mass star. We will assume 1\% of this total energy budget is carried by relativistic protons, $E_{\textup{p}} \sim 10^{51}$ erg. The total energy in muon neutrinos would then be $E_{\nu, \rm tot} = (1/8) E_{\rm p} \sim 10^{50}$ erg for efficient optically-thick pion production, after accounting for the pion decay yield and subsequent neutrino flavour oscillations. Convolving this implied energy $E_{\nu, \rm tot}$ with the effective area, $A_{\rm eff}$, of IceCube's high-energy neutrino alert selection\cite{2019ICRC...36.1021B}, we estimate the expected number of neutrino alerts. Approximating the sharply-peaked p$\gamma$ neutrino spectrum as a monoenergetic flux anywhere between  0.2 PeV $\lesssim \epsilon_\nu \lesssim 1$ PeV, we find $N_\nu = (E_{\nu, \rm tot}$ /$\epsilon_\nu)(A_{\rm eff}$ / 4$\pi D_{\rm L}^{2})  \sim 0.03$. Thus any optically-thick p$\gamma$ scenario would be sufficient to produce the neutrino under these assumptions. 

In contrast to a peaked p$\gamma$ neutrino spectrum, for pp production the neutrinos would instead follow a power law. Many of these neutrinos would then fall below the threshold of IceCube's alert selection. The associated gamma rays would however fall within the sensitive range of gamma-ray telescopes, so this scenario could be securely identified through a joint neutrino-gamma ray signal. While no gamma-ray emission was measured using the \textit{Fermi}-LAT telescope for AT2019dsg (see Methods), gamma-ray Cherenkov telescopes may be sensitive to the expected gamma-ray signal, and the corresponding low-energy (TeV) neutrino emission could confirm a hadronic origin. Conversely, the high optical depth of the UV photosphere would absorb any gamma rays accompanying p$\gamma$ neutrino emission\cite{2016PhRvD..93h3005W}. Some contribution from gamma-dark sources is required to explain the large astrophysical neutrino flux\cite{2016PhRvL.116g1101M}.

Given the different neutrino spectrum expectations, a search for accompanying lower-energy neutrinos could be used to probe the conditions at the site of proton interaction. IceCube has already searched for correlations between a sample of TDEs and a neutrino dataset dominated by lower-energy events, and reported that thermal TDEs account for less than 39\% of the diffuse astrophysical flux under the assumption of standard candles following a power-law spectrum\cite{2019ICRC...36.1016S}. The detection of a single TDE-neutrino association with our program would imply that at least $3\%$ of the astrophysical neutrino alerts arise from the TDE population, fully compatible with these pre-existing limits (see SI). 

As TDE discovery rates have increased substantially since the previous IceCube analysis\cite{2020arXiv200101409V, 2019ICRC...36.1016S}, future searches will be able to study neutrino emission from TDEs with much greater sensitivity. A measurement of O($\sim$1-10) TeV neutrinos without accompanying gamma rays would indicate that neutrino production is occurring in the X-ray photosphere, rather than in the UV photosphere. Indeed, such a detection would confirm the presence of a hidden X-ray source in the first place, while our electromagnetic observations cannot. Conversely, a lack of complementary low-energy neutrinos and gamma rays implies that only UV photons serve as a target. Neutrinos can uniquely serve as probes of the inner region of TDEs, using this novel method of extragalactic neutrino tomography. Now that a persistent central engine has been revealed in coincidence with a high-energy neutrino, we can begin to shed light on the role of TDEs as astrophysical accelerators.

%Indeed, the ZTF survey itself has only began science operations in March 2018, and the first ZTF high-energy neutrino follow-up was performed in May 2019\@. If AT2019dsg forms part of a neutrino-emitting TDE population, further such TDE-neutrino coincidence may thus be expected in the near future.

% Furthermore, analogous to the Greisen-Zatsepin–Kuzmin mechanism for suppression of ultra-high-energy cosmic rays, a photosphere model would also imply that TDEs could only accelerate protons to energies of $\sim$10 - 100 PeV, beyond which the blackbody cutoff would suppress ultra-high energy cosmic rays.

%\bibliography{ref.bib}

\newpage

{\bf References}\\[-5pt]

% \begin{itemize}

\section*{Correspondence} Correspondence and requests for materials
should be addressed to Marek Kowalski (email: \\ marek.kowalski@desy.de), Robert Stein~(email: robert.stein@desy.de) and Sjoert van Velzen~(email: sjoert@nyu.edu).

\section*{Acknowledgements} We thank Cecilia Lunardini, Andrew MacFadyen, Brian Metzger, Andrew Mummery, Alex Pizzuto, Nick Stone, Andrew Taylor and Walter Winter for fruitful discussions. We also thank the IceCube Collaboration for publishing high-energy neutrino alerts. We thank Seth Digel, Ke Fang, Deirdre Horan, Matthew Kerr, Vaidehi Paliya and Judith Racusin for useful feedback provided during Fermi collaboration review. R.S. is grateful to NYU for generously facilitating a visit to develop this work.  M.K. is grateful for the hospitality received from Columbia University and NYU during a sabbatical visit. 

This work was supported by the Initiative and Networking Fund of the Helmholtz Association through the Young Investigator Group program (A.F.).

S.v.V is supported by the James Arthur Postdoctoral Fellowship.

This research was partially supported by the Australian Government through the Australian Research Council's Discovery Projects funding scheme (project DP200102471). 

The work of M.B. is supported through the South African Research Chair Initiative of the National 
Research Foundation and the Department of Science and Innovation of South Africa, under SARChI Chair grant No. 64789. Any opinion, finding and conclusion or recommendation expressed in 
this material is that of the authors and the NRF does not accept any liability in this regard.

A.H. acknowledges support by the I-Core Program of the Planning and Budgeting Committee
and the Israel Science Foundation.

The UCSC transient team is supported in part by NSF grant AST-1518052, NASA/{\it Swift} grant 80NSSC19K1386, the Gordon \& Betty Moore Foundation, the Heising-Simons Foundation, and by a fellowship from the David and Lucile Packard Foundation to R.J.F. 

V.Z.G. is a Moore-Sloan, WRF Innovation in Data Science, and DIRAC Fellow.

A.G.Y.’s research is supported by the EU via ERC grant No. 725161, the ISF GW excellence center, an IMOS space infrastructure grant and BSF/Transformative and GIF grants, as well as The Benoziyo Endowment Fund for the Advancement of Science, the Deloro Institute for Advanced Research in Space and Optics, The Veronika A. Rabl Physics Discretionary Fund, Paul and Tina Gardner, Yeda-Sela and the WIS-CIT joint research grant;  AGY is the recipient of the Helen and Martin Kimmel Award for Innovative Investigation.

M.R. has received funding from the European Research Council (ERC) under the European Union’s Horizon 2020 research and innovation program (grant agreement n$^\circ$759194 - USNAC). 

The work of D.S. was carried out at the Jet Propulsion Laboratory, California Institute of Technology, under a contract with NASA.

The work of S.R. was supported by the Helmholtz Weizmann Research School on Multimessenger Astronomy, funded through the Initiative and Networking Fund of the Helmholtz Association, DESY, the Weizmann Institute, the Humboldt University of Berlin, and the University of Potsdam.

Based on observations obtained with the Samuel Oschin Telescope 48-inch and the 60-inch Telescope at the Palomar Observatory as part of the Zwicky Transient Facility project. ZTF is supported by the National Science Foundation under Grant No. AST-1440341 and a collaboration including Caltech, IPAC, the Weizmann Institute for Science, the Oskar Klein Center at Stockholm University, the University of Maryland, the University of Washington, Deutsches Elektronen-Synchrotron and Humboldt University, Los Alamos National Laboratories, the TANGO Consortium of Taiwan, the University of Wisconsin at Milwaukee, and Lawrence Berkeley National Laboratories. Operations are conducted by COO, IPAC, and UW. SED Machine is based upon work supported by the National Science Foundation under Grant No. 1106171. The National Radio Astronomy Observatory is a facility of the National Science Foundation operated under cooperative agreement by Associated Universities, Inc. The MeerKAT telescope is operated by  the South African Radio Astronomy Observatory, which is a facility  of the National Research Foundation, an agency of the Department of Science and Innovation.

This work was supported by the GROWTH (Global Relay of Observatories Watching Transients Happen) project funded by the National Science Foundation Partnership in International Research and Education program under Grant No 1545949. GROWTH is a collaborative project between California Institute of Technology (USA), Pomona College (USA), San Diego State University (USA), Los Alamos National Laboratory (USA), University of Maryland College Park (USA), University of Wisconsin Milwaukee (USA), Tokyo Institute of Technology (Japan), National Central University (Taiwan), Indian Institute of Astrophysics (India), Inter-University Center for Astronomy and Astrophysics (India), Weizmann Institute of Science (Israel), The Oskar Klein Centre at Stockholm University (Sweden), Humboldt University (Germany).

The Liverpool Telescope is operated on the island of La Palma by Liverpool John Moores University in the Spanish Observatorio del Roque de los Muchachos of the Instituto de Astrofisica de Canarias with financial support from the UK Science and Technology Facilities Council

Research at Lick Observatory is partially supported by a generous gift from Google.

The \textit{Fermi}-LAT Collaboration acknowledges support for LAT development, operation and data analysis from NASA and DOE (United States), CEA/Irfu and IN2P3/CNRS (France), ASI and INFN (Italy), MEXT, KEK, and JAXA (Japan), and the K.A.~Wallenberg Foundation, the Swedish Research Council and the National Space Board (Sweden). Science analysis support in the operations phase from INAF (Italy) and CNES (France) is also gratefully acknowledged. This work performed in part under DOE Contract DE-AC02-76SF00515. 

\section*{Author Contributions} R.S. first identified AT2019dsg as a candidate neutrino source, performed the neutrino analysis and was the primary author of the manuscript. M.K., R.S., and S.v.V. developed the multi-zone model. G.F., M.K. and R.S. performed the neutrino modelling. A.F, J.Ne., R.S. and S.R. scheduled and analysed ZTF ToO observations. J.C.A.M.J. and S.v.V. contributed the VLA observations. A.H., R.F., and I.S. contributed the AMI-LA observations. M.Bi., M.Bo., R.F., J.C.A.M.J. and P.W. contributed the MeerKAT observations. S.Ge. and S.v.V. requested and reduced the Swift-UVOT data. D.A.P. and K.T. contributed the LT observations. S.B.C., S.F. and S.Ge. performed X-rays observations and data analysis. S.Ga. analysed Fermi gamma-ray data. S.R. and S.v.V. analysed the ZTF data. J.B., E.C.B., R.B., S.B.C., V.C., M.F. V.Z.G., A.G., M.J.G, G.H., M.M.K., T.K., R.R.L., A.A.M, F.J.M., H.R., B.R., D.L.S., and M.T.S contributed to the implementation of ZTF. T.A, I.A, M.W.C, M.M.K and L.P.S enabled ZTF ToO observations. A.D., R.J.F., M.J.G., S.Ge., E.H., T.H., M.M.K, C.D.K., M.R., C.R.B., D.S, C.W., and Y.Y contributed to spectroscopic observations and data reduction. R.S. developed the ToO analysis pipeline. V.B., J.No. and J.v.S. developed Ampel, and contributed to the ToO analysis infrastructure. A.G.Y., A.K.H.K., and J.S. contributed to the manuscript and discussions. All authors reviewed the contents of the manuscript.

% \end{itemize}

\newpage

\newpage

\newpage

\clearpage
\newpage

\begin{methods}
\section*{ZTF Neutrino Follow-up}
 
 ZTF routinely images the visible Northern Sky once every three nights to a median depth of $20.5^{m}$, as part of a public survey\cite{2019PASP..131a8002B, 2019PASP..131g8001G}. For our neutrino follow-up program, this wide-field cadence is supplemented by dedicated Target-of-Opportunity (ToO) observations scheduled through the GROWTH ToO Marshal\cite{2019PASP..131d8001C}. 

 With ZTF, we have followed up eight neutrinos in the period from survey start on 2018 March 20 to 2020 March 31, out of a total of 31 neutrino alerts published by IceCube (see SI, Table S1). From 2019 June 17, IceCube published neutrino alerts with improved selection criteria to provide an elevated alert rate\cite{2019ICRC...36.1021B}. In addition to 1 of the 12 alerts under the old selection, ZTF  followed up 7 of the 19 alerts published under the new selection. In general, we aim to follow all well-localised neutrinos of likely astrophysical origin reported by IceCube which are visible to ZTF and can be observed promptly. Those alerts not observed by ZTF are summarised in the SI (Table S2). Of those 23 alerts not followed up by ZTF, the primary reasons were proximity to the Sun (8/23), alerts with poor localisation and low astrophysical probability (6/23) and alert retraction (4/23). For events which were reported with estimates of astrophysical probability, we chose not to follow up those that had both low astrophysical probability ($<$ 50\%) and large localisation regions ($>$ 10 sq.\,deg.). This astrophysical probability was not reported for high-energy starting events (HESE) under the old IceCube alert selection, nor for one recent alert, IC200107A, that was identified outside of the standard alert criteria\cite{stein:gcn26655}.

 Each neutrino localisation region can typically be covered by one or two ZTF observation fields. Multiple observations are scheduled for each field, with both $g$ and $r$ filters, and a separation of at least 15 minutes between images. These observations typically last for 300~s, with a typical limiting magnitude of $21.0^{m}$.  ToO observations are typically conducted on the first two nights following a neutrino alert, before swapping to serendipitous coverage as part of the public survey. Following observations, images are processed by IPAC\cite{2019PASP..131a8003M}, and alert packets are generated for significant detections from difference images\cite{2019PASP..131a8001P}.

 This alert stream of significant detections is then filtered by our follow-up pipeline built within the AMPEL framework\cite{ampel_followup_pipeline}, a platform for realtime analysis of multi-messenger astronomy data\cite{2019A&A...631A.147N}. Our selection is based on an algorithm for identifying extragalactic transients\cite{2019A&A...631A.147N}. We search ZTF data both preceding and following the arrival of the neutrino. In order to identify candidate counterparts to the neutrino, we apply the following cuts to ToO and survey data:

 \begin{itemize}
 	\item We reject likely subtraction artifacts using machine learning classification and morphology cuts\cite{2019PASP..131c8002M}.
 	\item We reject moving objects through matches to known nearby solar system objects\cite{2019PASP..131a8003M}. We further reject moving objects by requiring multiple  detections for each candidate (i.e, at the same location) separated temporally by at least 15 minutes.
 	\item We remove stellar sources by rejecting detections cross-matched\cite{2018PASP..130g5002S} to objects with measured parallax in \textit{GAIA} DR2 data\cite{2018A&A...616A...1G}, defined as non-zero parallax with a significance of at least 3$\sigma$. We further reject likely stars with machine learning classifications\cite{2018PASP..130l8001T}, based on sources detected by Pan-STARRS1\cite{2016arXiv161205560C}, removing those objects with an estimated stellar probability greater than 80\%.
 	\item We identify likely AGN by cross-matching to the WISE survey and applying IR color cuts\cite{2010AJ....140.1868W}. We reject detections consistent with low-level AGN variability. 
 	\item We require that objects lie within the reported 90\% error region to ensure spatial coincidence, and that they are detected at least once following the neutrino arrival time to ensure temporal coincidence.
 \end{itemize}

 These cuts typically yield $\sim$0.2 candidates per square degree of sky. Promising candidates are prioritised for spectroscopic classification, to confirm or rule out a possible association with a given neutrino. 

 AT2019dsg (R.A.[J2000] = 314.26 deg, Dec[J2000] = +14.20 deg) was spatially-coincident with the 90\% localisation of the neutrino IC191001A\cite{stein:gcn25913} (R.A. = 314.08$^{+6.56}_{-2.26}$ deg,  Dec = +12.94$^{+1.50}_{-1.47}$ deg), at a distance of 1.27 deg to the best-fit position. It was also temporally coincident, being detected by ZTF in our ToO observations following the neutrino detection. There were additionally three candidate supernovae found in the error region of IC191001A,  consistent with background expectations.  AT2019dsg was the first TDE identified by our pipeline, and the first TDE to be reported in coincidence with any high-energy neutrino.

\section*{Probability of Chance Coincidence}
 
 % TDEs are rare transients\cite{2018ApJ...852...72V}, with an estimated local rate of $R(z=0.1) = (8 \pm 4) \times 10^{-7} \; {\rm Mpc}^{-3}\; {\rm yr}^{-1}$. 
 During the first 18 months of survey operations, ZTF identified 17 TDEs\cite{2020arXiv200101409V}, distributed over 28000 deg of observed sky (the ZTF survey footprint, after removing sources with a Galactic latitude $|b|<7$). Of these TDEs, each was typically detected for $\sim$6 months\cite{2020arXiv200101409V}. We thus estimate that the density of ZTF-detected TDEs is approximately 2.0 $\times 10^{-4}$ per sq.\,deg. of sky in the survey footprint at any given time. Our follow-up pipeline requires that any candidate be detected by ZTF in ToO observations following a neutrino, in order to establish temporal coincidence. We assume that our neutrino pipeline does not have a significantly higher selection efficiency than the dedicated ZTF program to identify TDEs\cite{2020arXiv200101409V}, and thus that the latter provides a reasonable estimate on the background rate of TDEs passing our pipeline.

 Those TDEs with radio detections are considered the most promising candidates for neutrino production, as the radio emission serves as a tracer for the particle acceleration required in neutrino sources. We can consider the fraction of TDEs which would additionally be detected in radio, assuming that all could be observed. Among the ZTF sample of confirmed TDEs, we undertook radio follow-up observations with the VLA for 6, of which 2 were detected. Taking this implied radio-emitting fraction of 33\%, we then find a final density of 5.9 $\times 10^{-5}$ radio-emitting TDEs per sq.\,deg. of surveyed sky. 

 ZTF has followed-up eight neutrinos up to January 2020, and has covered a combined localisation region of 81.05 sq.\,deg (see SI, Table S1). With this sky area, the expected number of coincident radio-detected TDEs across all of our neutrino follow-up campaigns is 4.8 $\times 10^{-3}$. The Poisson probability of observing at least one radio-emitting TDE during our entire neutrino follow-up campaign is thus 4.8$ \times 10^{-3}$. 

 As radio follow-up observations of ZTF TDEs were biased towards those most likely to be detectable, this estimate is an overly conservative one. Because the bolometric energy flux derived from UV/optical observations (i.e., the blackbody luminosity over the square of the distance) serves as a proxy the non-thermal emission, TDEs which were bright under this metric were preferentially selected for radio observations. To avoid this selection bias, we can instead directly use this bolometric energy flux  as a proxy for neutrino flux to identify the most promising candidates for neutrino detection, namely those TDEs which are both nearby and luminous. Of the 17 TDEs observed by ZTF, AT2019dsg ranks second in this metric. The probability of finding a TDE in our neutrino follow-up campaign with a bolometric energy flux that is at least as high as AT2019dsg is thus 1.9$ \times 10^{-3}$. 

 Like most other studies in neutrino astronomy\cite{2018Sci...361.1378I, 2020PhRvL.124e1103A}, these chance coincidence probability estimates do not account for the so-called ``look-elsewhere effect" from multiple possible hypotheses. In our case, the ZTF program has sensitivity to four theoretically-motivated neutrino population hypotheses (TDEs, core-collapse supernovae, gamma-ray bursts and AGN flares)\cite{2019PASP..131g8001G}. The impact of the testing multiple hypotheses is thus modest, and a chance coincidence explanation for AT2019dsg-IC191001A remains unlikely.

 Of these four, it should also be noted that TDEs are the one to which our program is most sensitive. As detailed in the SI, follow-up programs are generally most effective in identifying neutrino emission from TDEs, since this flux should be dominated by nearby sources which can be detected by telescopes such as ZTF. Moreover, given their low rate, any individual TDE-neutrino association will be easier to identify than for more abundant populations such as AGN or supernovae. TDEs also evolve sufficiently slowly to enable extensive photometric and spectroscopic follow-up, in marked contrast to fast transients such as GRB afterglows, leading to a higher detection efficiency.

 \section*{Optical/UV Observations}
 
 Prior to the detection of IC191001A, AT2019dsg had already been repeatedly detected by ZTF P48 telescope as part of the public MSIP survey, most recently on 2019 September 28. These data were supplemented by photometric observations from the 2m Liverpool Telescope\cite{2004SPIE.5489..679S} and SEDM\cite{Blagorodnova18,Rigault19} photometry\cite{fst+2016} obtained using the P60 telescope on Mt Palomar. ToO observations of the neutrino localisation field began on 2019 October 1, 7.4 hours after the neutrino detection. A second set of observations were performed the following night. In all of these images AT2019dsg was clearly visible. 

 UV observations of AT2019dsg were conducted as part of a systematic survey of UV properties of all ZTF-identified TDEs\cite{2018ApJ...852...72V}, using the UltraViolet/Optical Telescope\cite{2005SSRv..120...95R} (UVOT) on board the \textit{Neil Gehrels Swift Observatory} (\textit{Swift})\cite{2004ApJ...611.1005G}. Data were reduced with \texttt{uvotsource} using an aperture of 7" to capture the entire galaxy (the host flux density was subtracted based on the best-fit galaxy model\cite{2018ApJ...852...72V} and uncertainties on this baseline are propagated into the reported UVOT difference photometry). The first UV observation was performed 15 days after the optical peak on 2019 May 17, and a bright source spatially coincident with the TDE was detected. Subsequent observations continued at a cadence of 2--3 days, up to 2019 September 7. In this period, AT2019dsg continued to steadily dim. An additional observation occurred shortly before the neutrino detection on 2019 September 27. Follow-up observations were then triggered by the identification of a possible association with IC191001A\cite{2019ATel13160....1S}, beginning on 2019 October 5. 

 The optical/UV data are summarised in the SI (Table S3). We note that in the final ZTF observations, the source appears to redden in the optical bands. This could be a signature of reverberation due emission from dust heated by the TDE\cite{2016ApJ...829...19V,2016MNRAS.458..575L}; this dust can reach a temperature of $\sim 2000$ K. An important caveat is that the contrast between the transient emission and the host is very small for these late-time optical detections, so the residuals in the difference image may need to be corrected to account for small systematic offsets. That can only be investigated when the images for this portion of the public survey are published as part of the next ZTF data release. We note that the UV observations are not subject to the same uncertainty because even at late times the transient UV flux is about an order of magnitude brighter than the host baseline.

 %\subsection*{Spectroscopy.}
 AT2019dsg was first classified as a TDE by ePESSTO+ on on 2019 May 13\cite{2019ATel12752....1N}, and the redshift of AT2019dsg was measured to be $z=0.051$. Further high-resolution spectroscopic observations were conducted using the De Veny Spectrograph on the 4.3m Lowell Discovery Telescope (LDT, PI: Gezari), the Kast Double Spectrograph on the 3m Lick Observatory Shane Telescope (Lick, PI: Foley)\cite{Miller93}, and the Low Resolution Imaging Spectrograph on the 10m Keck Telescope (Keck, PI: Graham)\cite{Oke95}, with the most recent spectrum on 2019 September 25. These spectra confirm that AT2019dsg belongs to the common spectroscopic class of TDEs with Bowen fluorescence emission lines and broad H$\alpha$ emission lines\cite{2020arXiv200101409V}. We note that the Ca triplet is also clearly visible in our late-time spectra (rest-frame 8498 \AA, 8542\AA ~and 8662 \AA), so the SMBH mass could in principle be inferred more precisely using higher-resolution spectroscopy of this feature\cite{2005MNRAS.359..765G}. Following the identification of AT2019dsg as a candidate neutrino source, additional high-resolution spectra of the source were taken with the 200in Hale Telescope Double Spectrograph at Palomar Observatory (P200, PI: Kasliwal \& Kulkarni) on 2019 October 3 and again with  Lick on 2019 October 5 and 2019 October 29 (shown in Extended Data Figure 1). There is no evidence of any significant spectral evolution between these spectra and the most recent pre-neutrino spectrum from 2019 September 25, and the spectral evolution of AT2019dsg is consistent with that of other TDEs\cite{2020arXiv200101409V}. 

 \section*{X-ray Observations}

 AT2019dsg was first observed in X-rays on 2019 May 17 by the X-Ray Telescope (XRT)\cite{2005SSRv..120..165B}, also on board \textit{Swift}\cite{2004ApJ...611.1005G}, as part of a program to categorise the X-ray properties of TDEs. AT2019dsg was detected at high significance at this epoch, with a measured energy flux of $F_{X}\sim 4 \times 10^{-12}$ erg cm$^{-2}$ s$^{-1}$ (0.3--10 keV). Observations continued with a cadence of 2--3 days, and indicated a sharply-declining X-ray flux (see SI, Table S4). The source was last detected on 2019 June 14, and not detected again in any of the following observations continuing until 2019 September 7. An additional observation was performed with the \textit{X-ray Multi-Mirror Mission} (\textit{XMM-Newton}) telescope on 2019 May 30, in the range 0.3-10 keV. The \textit{XMM-Newton} EPIC-pn observations (programs 082204 and 08425; P.I. Gezari) were taken in Wide window Thin1 filter mode and reduced using standard techniques with the \textit{XMM-Newton}\cite{2001A&A...365L...1J} Science Analysis System (SAS). The source extraction region was a circle of radius 35 arcsec at the location of the optical transient in the X-ray image, and the background was measured using a 108-arcsec circular region (shown in Extended Data Figure 2). The \textit{XMM} spectrum was binned using the \texttt{GRPPHA} command, such that there were at least 20 counts contained in each bin. It was then fit ($\chi^2/\rm{dof}=59.26/65$) with the  disk blackbody (\texttt{diskbb}) model with Galactic\cite{HI4PI2016} and intrinsic ($N_{\rm{H}} \sim 4 \times 10^{20}$ cm$^{-2}$) absorption described using the \texttt{phabs} model in \texttt{XSPEC} v12.9.1\cite{1996ASPC..101...17A}. The flux was consistent with those of \textit{Swift}-XRT, and provided a high signal-to-noise X-ray spectrum well-fitted with a single disk temperature of $T_{\rm disk} = 10^{5.9}$ K (0.072 $\pm 0.005$ keV), shown in Extended Data Figure 3. Following the identification of AT2019dsg as a candidate counterpart to IC191001A\cite{2019ATel13160....1S}, additional X-ray observations were triggered. AT2019dsg was again not detected, with the first \textit{Swift}-XRT observation occurring on 2019 October 5. An additional \textit{XMM} observation on 2019 October 23 yielded a deep upper limit of $9 \times 10^{-14}$ erg cm$^{-2}$ s$^{-1}$ (0.3--10 keV) using the same thermal model, computed at the 3$\sigma$ confidence level using the \textit{XMM} SAS/HEASARC command \texttt{eregionanalyse}. 

 \section*{Radio Observations}

 % VLA paragraphs added by James Miller-Jones
 Four observations of AT2019dsg were performed with the Karl G.\ Jansky Very Large Array (VLA) under project code 19A-395 (PI: van~Velzen), on 2019 May 22, June 19, August 8 and October 5.  The array was in its moderately-extended B configuration (maximum baseline 11\,km) for the first two epochs, and in its most extended A-configuration (maximum baseline 36\,km) for the final two epochs.  Our first epoch, on May 22, was a detection experiment, and we observed only in the 8--12\,GHz band.  Having established the presence of radio emission, we observed over a broader range of frequencies in the subsequent three epochs, using the 2--4\,GHz, 4--8\,GHz, and 8--12\,GHz bands.  We used 3C\,48 as a bandpass and flux density calibrator on May 22, and 3C\,286 for the other three epochs.  We used the nearby extragalactic sources ICRF J204945.8+100314 
 % MFB: we should give source names that resolve on e.g. Simbad, and good to mention its an ICRF source - our posn accuracy should be good.
 (at 4--8 and 8--12\,GHz) and ICRF J203533.9+185705 (at 2--4\,GHz) to determine the complex gain solutions, which were interpolated to AT2019dsg. We used the VLA pipeline to perform external gain calibration, and after removing residual radio frequency interference, we imaged the data within CASA, using Briggs weighting with a robust parameter of 1.  We split each baseband into multiple frequency bins for imaging (1\,GHz bins above 4\,GHz, and 0.5\,GHz bins below that) to provide better sampling of the broadband spectrum, allowing more precise constraints on the turnover frequency, and better spectral modelling.

 % two paragraphs from AMI 
 Radio observations of the field of AT2019dsg were also conducted using the AMI Large Array (AMI-LA)\cite{2008MNRAS.391.1545Z,2018MNRAS.475.5677H}. AMI-LA is a radio interferometer comprised of eight 12.8m-diameter antennas producing 28 baselines that range from 18m up to 110m, which operates with a 5 GHz bandwidth around a central frequency of 15.5 GHz. We observed AT2019dsg on several epochs (see SI, Table S5) for four hours each. Initial data reduction, editing, and calibration of the phase, and flux density, was carried out using \verb reduce_dc , a customized AMI data reduction software package\cite{2015MNRAS.453.1396P}. Phase calibration was conducted using short interleaved observations of ICRF J205135.5+174336,
 %MFB was: J2051+1743, 
 while for absolute flux density calibration we used 3C\,286. Additional flagging and imaging were performed using Common Astronomy Software Application (CASA)\cite{McMullin2007}. All of our observations showed a source consistent with the location of AT2019dsg. We used the CASA task IMFIT to find the source flux and position. 

 Further observations of AT2019dsg were conducted with the South African MeerKAT telescope, on 2019 June 19, July 29, October  5, and November 30, with each session being $\sim$2~h long.  We used ICRF J193925.0-634245 as a flux-density calibrator, and ICRF J213032.8+050217 as a phase and amplitude calibrator.  The initial calibration was done using the IDIA MeerKAT pipeline (\url{https://idia-pipelines.github.io/docs/processMeerKAT}), which is implemented in CASA\@.  The observed band was 860 MHz wide and centred on 1280 MHz. We imaged the whole primary beam ($\sim 1$\arcdeg) using the CLEAN algorithm (CASA: tclean) in order to remove sidelobes from the many (unrelated) sources within the primary beam.  The total CLEAN flux density in the field was $\sim$1~Jy, and the peak brightness in the images was about 48 mJy~beam$^{-1}$ (not related to AT2019dsg). Since residual small calibration errors dominated the image rms background in the initial images, we self-calibrated the data in both phase and amplitude, with the mean amplitude gain being fixed at unity to minimise any drifting of the flux-density scale. The resolution is slightly different in each epoch, but was $\sim$11\arcsec\ north-south, and $\sim$6\arcsec\ east-west.  Image rms background levels also varied, ranging between 25 and 32 $\mu$Jy~beam$^{-1}$.  There was no sign of extended emission or confusing sources near AT2019dsg. The flux density was determined by fitting an elliptical Gaussian with the same geometry as the restoring beam to the images.  

 The measured flux densities from our radio observations are reported in Extended Data Figure 7. For all radio observations, the reported uncertainties include both the image background rms and a 5\% fractional calibration uncertainty, added in quadrature.

 \section*{Gamma-Ray Observations}
 
 We analysed data from the %pair-conversion telescope 
 \textit{Fermi} Large Area Telescope (\textit{Fermi}-LAT)\cite{2009ApJ...697.1071A}, sensitive to gamma rays with energies from 20 MeV to greater than 300 GeV. During its sky-survey operations, the pair-conversion telescope \textit{Fermi}-LAT scans the entire sky about every three hours, and can monitor the variable gamma-ray sky over short and long timescales. We studied the region of AT2019dsg in three different time intervals, motivated by the multi-wavelength behavior of the source. The first interval (G1) includes 130 days of observations that include the peak of the optical emission from 2019 April 4 to 2019 August 12. The second one (G2) spans 2019 August 12 to 2019 November 20 and covers the apparent UV plateau and the peak of the radio emission. The third period (G3) integrates the whole period between the start of G1 up to 2020 January 31. We use the photon event class from Pass 8 \textit{Fermi}-LAT data (P8R3\_SOURCE), and select a $15\arcdeg \times 15\arcdeg$ Region of Interest (RoI) centered on the AT2019dsg position derived from optical observations, with photon energies from 100 MeV to 800 GeV. We use the corresponding LAT instrument response functions P8R3\_SOURCE\_V2 with the recommended spectral models \textit{gll\_iem\_v07.fits} and \textit{iso\_P8R3\_SOURCE\_V2\_v1.txt} for the Galactic diffuse and isotropic component respectively. To minimise contamination from gamma rays produced in the Earth's upper atmosphere, we require an instrumental zenith angle $\theta<90\arcdeg$ for all events, in addition to the standard data quality cuts suggested by the {\it Fermi} Science Support Center (\url{https://fermi.gsfc.nasa.gov/ssc/data/analysis/scitools/data_preparation.html}). We perform a likelihood analysis, binned spatially with $0.1 \arcdeg$ resolution and 10 logarithmically-spaced bins per energy decade, using the \textit{Fermi}-LAT ScienceTools package (fermitools v1.0.1) along with the \textit{fermipy} package v0.17.4\cite{2017ICRC...35..824W}.

 A search was already performed within the 90\%\ error region during both the 1-day and 1-month period prior to the arrival of the high-energy neutrino\cite{garrappa_buson:gcn25932}. No new gamma-ray source was identified, and there was no significant ($\geq 5 \sigma$) detection for any source from the fourth \textit{Fermi}-LAT point source catalog (4FGL\cite{2019arXiv190210045T}). Here, we specifically test a point-source hypothesis at the position of AT2019dsg under the assumption of a power-law spectrum. We find TS = 0 for all intervals, where TS is twice the difference in the maximum $\log \mathcal{L}$ of an ROI model with and without the source, where $\mathcal{L}$ is the likelihood of the data given the model. Upper limits for the energy flux (integrated over the whole analysis energy range) have been derived for a power-law spectrum ($dN/dE \propto E^{-\Gamma}$) with photon power-law index $\Gamma$ = 2 and are listed in Extended Data Figure 6, along with the respective time intervals.

 In all three time intervals, we detect a new, non-catalogued gamma-ray emitter in the RoI at a significance $\geq 5 \sigma$. This source lies just outside the IC191001A 90$\%$ error region, as indicated in Extended Data Figure 4. The source, which we label \textit{Fermi}-J2113.8+1120, is likely the gamma-ray counterpart of the radio-loud object GB6 J2113+1121, classified as a flat-spectrum radio quasar with redshift $z = 1.63$\cite{2013ApJ...767...14P}. The detection of an unrelated gamma-ray blazar within the neutrino uncertainty area is consistent with the background estimation. On average 1.5 4FGL gamma-ray blazars are expected in 20 sq.\,deg. A lightcurve analysis (Extended Data Figure 5) reveals that the source was last detected in gamma rays one month prior to the IC191001A detection, and was not significantly detected again until another month after the neutrino detection. These results are compatible with the findings of the realtime follow-up of the region, which found no evidence for emission on timescales of 1-day and 1-month prior the detection of IC191001A \cite{2019GCN.25932....1G}.

 Such a long apparent lag between gamma-ray emission and neutrino emission is disfavored by recent studies on the temporal behavior of hadronic processes in blazars\cite{2015ApJ...802..133D,2019NatAs...3...88G}, suggesting that the blazar is unlikely to have produced the neutrino. %for which a shared origin of GeV photons and high-energy neutrinos from the same processes is disfavoured by recent studies on the temporal behavior of hadronic processes in blazars\cite{2015ApJ...802..133D,2019NatAs...3...88G}. 
 Hence, given the lack of any obvious connection between the gamma-ray observations of \textit{Fermi}-J2113.8+1120 and IC191001A, we do not discuss this source any further.

 The HAWC observatory also reported a search for transient gamma-ray emission on short timescales in the localisation of IC191001A\cite{ayala:gcn25936}, and set a limit for their most significant position at 95\% confidence of $E^{2} dN/dE = 3.51 \times 10^{-13} (\textup{E}/\textup{TeV})^{-0.3}$ TeV cm$^{-2}$ s$^{-1}$, in the energy range 300 GeV to 100 TeV, for the period from 2019 September 30 05:46:52 UTC to 2019 October 02 06:03:29 UTC. We note that this search covered a relatively large region of the sky, and thus had a large associated trial factor. A dedicated search at the position of AT2019dsg would be more sensitive, especially one that additionally targeted the longer period over which the central engine is active.

 \section*{Radio Analysis}
 
 The four radio spectral energy distributions (SEDs) for AT2019dsg can be described by synchrotron emission from a population of relativistic electrons. We assume that the electrons are accelerated into a power-law distribution in energy $dN_{e}/d\gamma \propto \gamma^{-p}$. We expect that the lowest-energy electrons emit their synchrotron radiation below the synchrotron-self absorption frequency with negligible free-free absorption, and in this case the shape of the radio SED is determined by just 3 free parameters, the electron power-law index $p$, the magnetic field $B$ and the source radius $R$: 
 \begin{equation}\label{eq:Fsync}
     F_{\nu,\rm sync}(t) \propto \frac{j_\nu(B(t), p)}{\kappa_\nu(B(t), p)} (1-e^{-\kappa_\nu R(t)})
 \end{equation}
 here $j_\nu$ and $\kappa_\nu$ are the emission and absorption coefficients, respectively. The normalization of Eq.~\ref{eq:Fsync} depends on the source geometry and the so-called microphysical parameters ($\epsilon_e$, $\epsilon_B$) which will be treated separately below.  Similar to the case of radio-emitting TDE ASASSN-14li\cite{2016Sci...351...62V,2016ApJ...819L..25A}, we might expect some steady radio emission from the host galaxy. This baseline flux density is parameterised as 
 \begin{equation}\label{eq:Fhost}
     F_{\nu,\rm baseline} = F_{\rm baseline} \left(\frac{\nu}{1.28 \textup{ GHz}}\right)^{\alpha_{\rm baseline}}
 \end{equation}
 such that the total flux density is given by 
 \begin{equation}\label{eq:Ftotal}
     F_{\nu,\rm total} = F_{\nu \rm, baseline} + F_{\nu,\rm sync} 
 \end{equation}
 The magnetic field and radius are allowed to change for each epoch, while $F_{\rm baseline}$, $\alpha_{\rm baseline}$ and $p$ are assumed to be constant during our radio observations. 

 Using Eq.~\ref{eq:Ftotal} to describe the synchrotron spectrum, we apply a Markov chain Monte Carlo approach\cite{Foreman-Mackey13} to determine a posterior probability distribution of the electron power-law index, as well as the peak frequency ($\nu_{\rm peak}$) and peak flux density ($F_{\rm peak}$) for each radio epoch (Extended Data Figure 8). We allow the measurement variance to be underestimated by some fractional amount $f$ (see, e.g., ref. \cite{Guillochon18}). 

 The last epoch of VLA observations, which has the best coverage of the optically thin part of the radio SED, yields $p=3.0 \pm 0.15$ and we use this as a Gaussian prior when fitting all data simultaneously.  We use a flat (uninformative) priors for the other parameters and we cap $\alpha_{\textup{baseline}}$ at 0 and $F_{\textup{baseline}}$ at 0.1 mJy (because the baseline flux density should be optically thin, $\alpha < 0$, and cannot exceed the observed post-TDE radio flux density). For the time-independent parameters we find: $F_{\textup{ baseline}}=0.09\pm 0.01$ mJy, $\alpha_{\textup{baseline}} = -0.2\pm 0.1$, $p = 2.9 \pm 0.1$, and $\ln f=-3.4$. We find no significant covariance between the baseline flux density parameters and the peak frequency or peak flux density. 

 To estimate the radius and energy of the radio-emitting region from the posterior distribution of $F_{\rm peak}$, $\nu_{\rm peak}$, and $p$, we use the scaling relations from {Barniol Duran}, {Nakar}, \& {Piran} (2013)\cite{2013ApJ...772...78B}. These relations depend on the electron power-law index and we propagate the uncertainty on $p$ into our estimates of $R_{\rm eq}$ and $E_{\rm eq}$. Additional assumptions for the geometry and the microphysical parameters are required. For the geometry of the outflow, our default model is two conical emitting regions with half-opening angles $\phi=$30~deg, which yield an area covering factor $f_A=1-\cos \phi=0.13$ and a volume factor $f_V=2/(3\tan \phi )=1.15$ (here we follow the convention\cite{2013ApJ...772...78B} that $f_A=1$ and $f_V=4/3$ parameterise a spherical outflow in the Newtonian limit). 

 The equipartition energy, $E_{\rm eq}$ is obtained under the assumption that the system contains only electrons and magnetic fields (both uniformly-distributed) and that the total energy is minimised for $E_{B} = (6/11)$ $E_{e}$. However we expect that protons carry the bulk of the energy and we parameterise this energy in protons by $\epsilon_e \equiv E_e/E_p$ with $E_p$ the total energy in relativistic protons.  After this adjustment, $E_{\rm eq}$ is increased by $(1+1/\epsilon_e)^{11/(13+2p)}$. Finally, for systems that are not in equipartition, the energy is larger by a factor\cite{2013ApJ...772...78B} $(11/17) \eta^{-6/17} + (6/17) \eta^{11/17}$, with $\eta\equiv (\epsilon_B/(1-\epsilon_B))/(6/11)$, with $\epsilon_B$ the fraction of total energy that is carried by the magnetic field. Motivated by observations of GRB afterglows\cite{Granot14,Fong15}, supernovae\cite{2013MNRAS.436.1258H} and the relativistic TDE Swift~J1644+57\cite{2018ApJ...854...86E}, we use $\epsilon_e=0.1$ and $\epsilon_B=10^{-3}$. %For these parameters, the equipartition energy and radius are scaled by factors $25$ and 0.78, respectively, see Table~\ref{tab:syncinf}. 

 From the equipartition magnetic field strength inferred from the first epoch of radio observation (see Extended Data Figure 8) we estimate that the cooling time of the electrons that emit at 10~GHz is 500~days. For the last epoch, the field strength has decreased by a factor of 3 and now the cooling time is an order of magnitude longer. We can thus conclude that, unless the magnetic field energy density is much higher than the equipartition value ($\epsilon_B/\epsilon_e\gg1$), the observed slope of the optically thin part of the radio SEDs reflects the intrinsic power-law index of the electrons. 

 From our synchrotron analysis we also obtain the number density of relativistic electrons, which in  turn yields a lower limit to the total particle density in the radio region. This estimate is relevant for the pp scenario of pion production, which requires a target density of at least $\sim 10^8$~cm$^{-3}$ to have sufficient optical depth. For the energy and radius of last radio epoch, which was obtained a few days after the neutrino detection, we find an electron density of $10^{3.4\pm 0.1}$~cm$^{-3}$ (see Extended Data Figure 8).

 We note that there is no evidence of correlation between the X-ray and radio emission in AT2019dsg, in contrast to coupling found for ASASSN-14li\cite{2018ApJ...856....1P}. Such a correlation would only be expected if the X-ray luminosity of AT2019dsg served as a tracer of disk power, but the rapid observed fading indicates that the observed X-ray emission in AT2019dsg is instead driven either by varying obscuration or temperature evolution. 

\newpage

{\bf Method References}\\[-5pt]

\newpage

\section*{Data availability} The data that support the plots within this paper, and other findings of this study are available from \href{https://github.com/robertdstein/at2019dsg}{https://github.com/robertdstein/at2019dsg}, and at DOI: \href{https://doi.org/10.5281/zenodo.4308124}{10.5281/zenodo.4308124}.
%Reduced optical spectra (Figure~\ref{fig:spectrum} will posted to TNS.  

\section*{Code availability} Python scripts used to perform significant calculations, and to reproduce all figures, are available from \href{https://github.com/robertdstein/at2019dsg}{https://github.com/robertdstein/at2019dsg}, and at DOI: \href{https://doi.org/10.5281/zenodo.4308124}{10.5281/zenodo.4308124}.

\section*{Competing Interests} The authors declare that they have no
competing financial interests.

\newpage

\renewcommand{\thefigure}{ED\arabic{figure}}
\renewcommand{\thetable}{ED\arabic{table}}
\setcounter{figure}{0}

\begin{figure*}
    \centering
    \includegraphics[width=\textwidth]{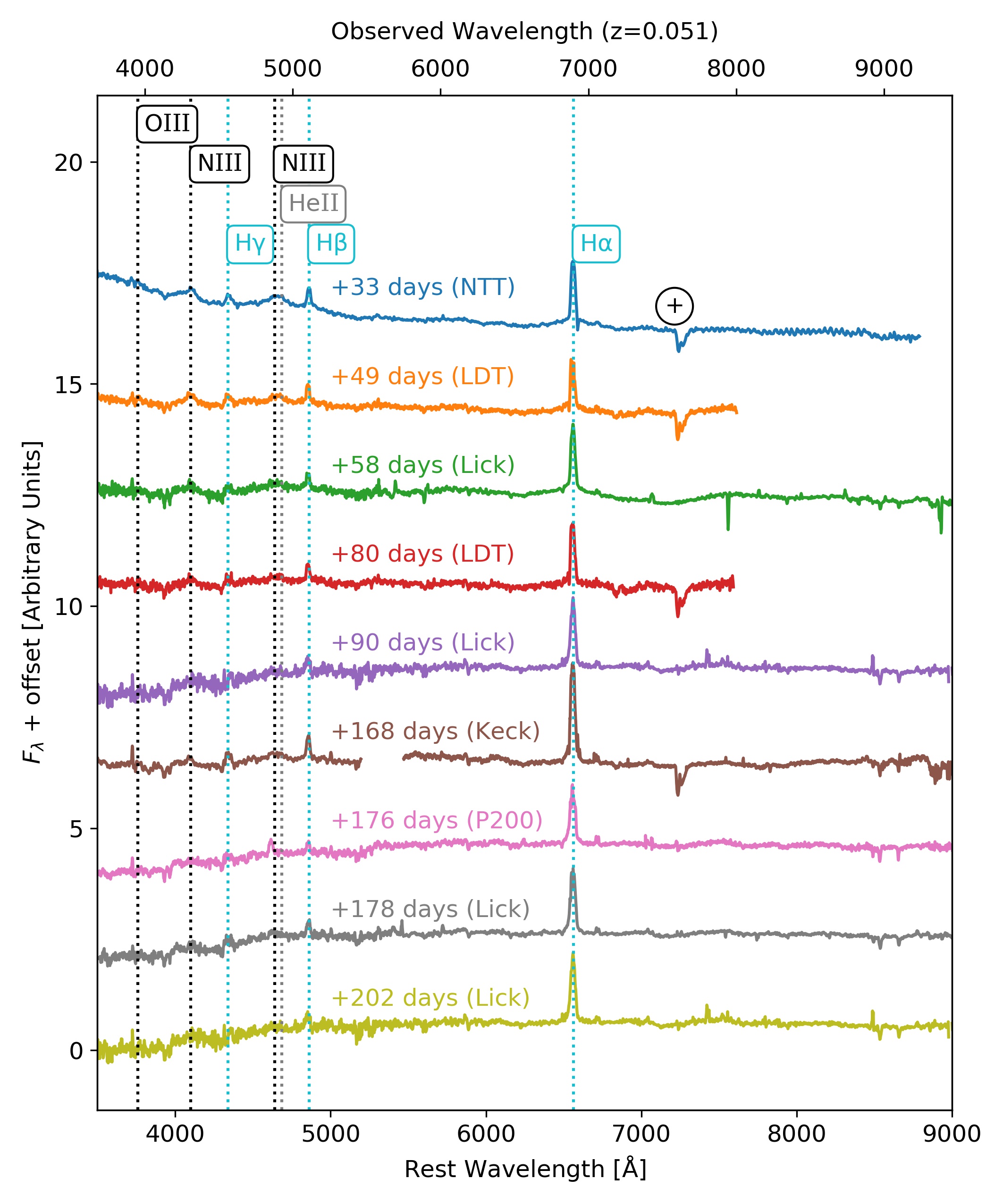}
    \caption{\textbf{Spectroscopic evolution of AT2019dsg}. The public classification spectrum taken with the NTT\cite{2019ATel12752....1N} is plotted in blue, along with later spectra from LDT, Lick, Keck and P200. Each is plotted in arbitrary units, offset for display purposes. The Balmer lines are highlighted in cyan, the HeII lines in gray, and the Bowen fluorescence lines (OIII at 3760\AA, NIII at 4100\AA ~and 4640\AA) in black. Telluric lines are marked with +.}
    \label{fig:spectrum}
\end{figure*}

\begin{figure*}
    \centering
    \includegraphics[width=\textwidth]{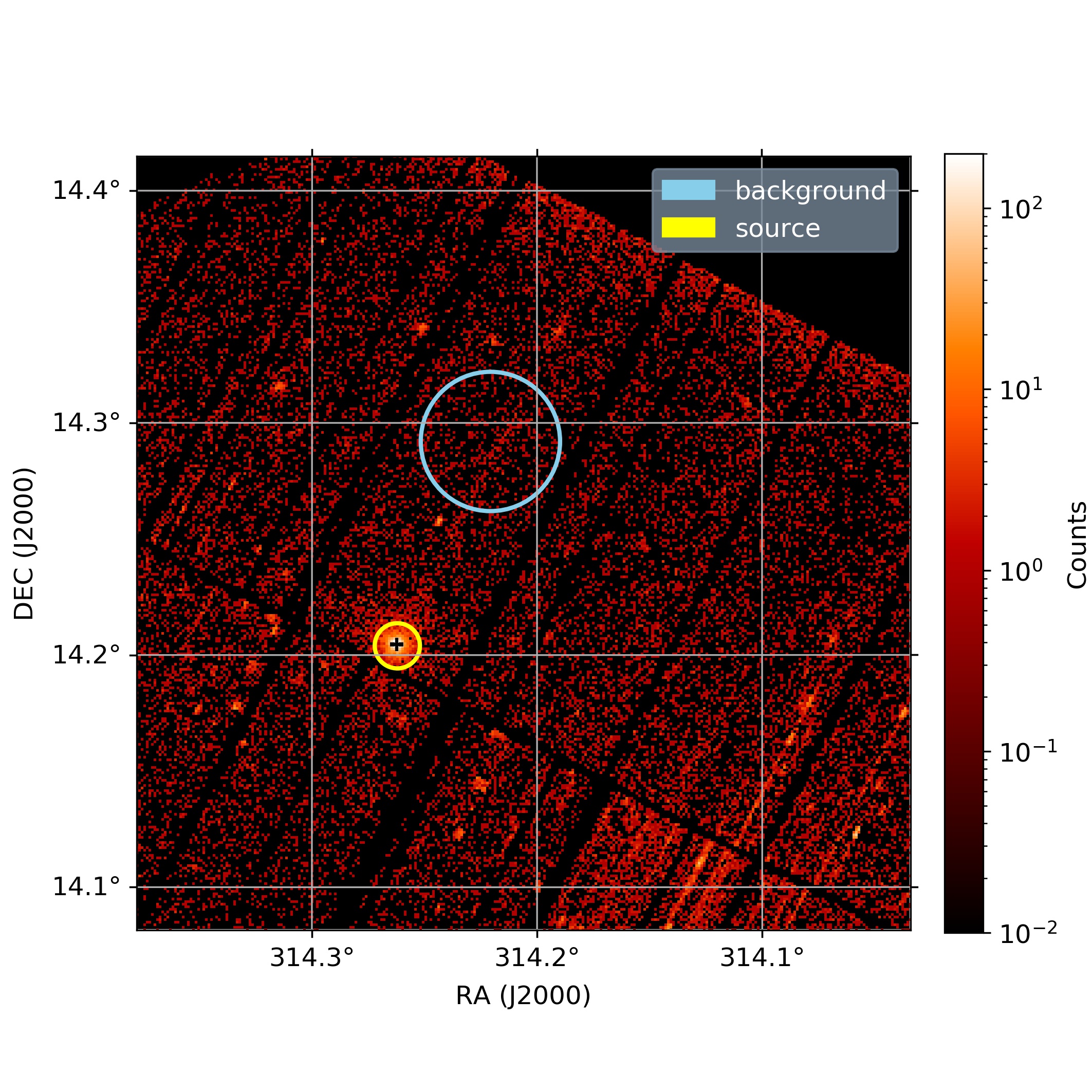}
    \caption{\textbf{X-ray count map from \textit{XMM-Newton}}. The image was taken 50 days after discovery. The green circle indicates the source region, while the red circular region was used to measure the background. The best-fit position derived from optical observations is spatially-coincident with the center of the X-ray source region.}
    \label{fig:xraymap}
\end{figure*}
\newpage

\begin{figure*}
    \centering
    \includegraphics{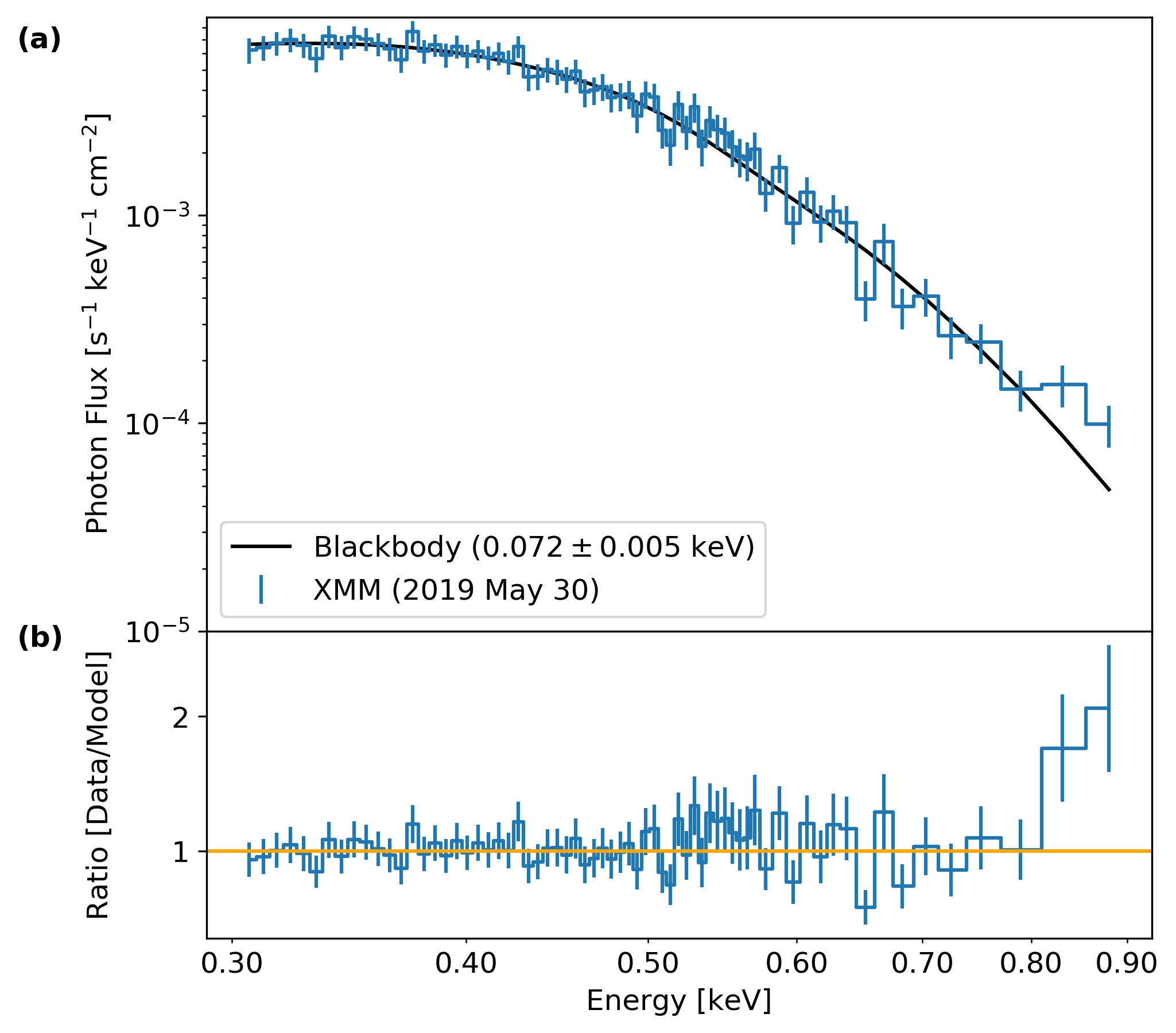}
    \caption{\textbf{Soft X-ray spectrum of AT2019dsg}. Panel (a) shows the photon flux as a function of energy in blue, from the XMM Newton spectrum taken on 2019 May 30. The spectrum was fitted with an absorbed disk blackbody model, shown in black. Panel (b) shows the ratio of the data to the model, with the horizontal orange line indicating unity. Error bars represent 1$\sigma$ intervals.}
    \label{fig:xrayspec}
\end{figure*}
\newpage

\begin{figure}
    \centering
    \includegraphics[width=\textwidth]{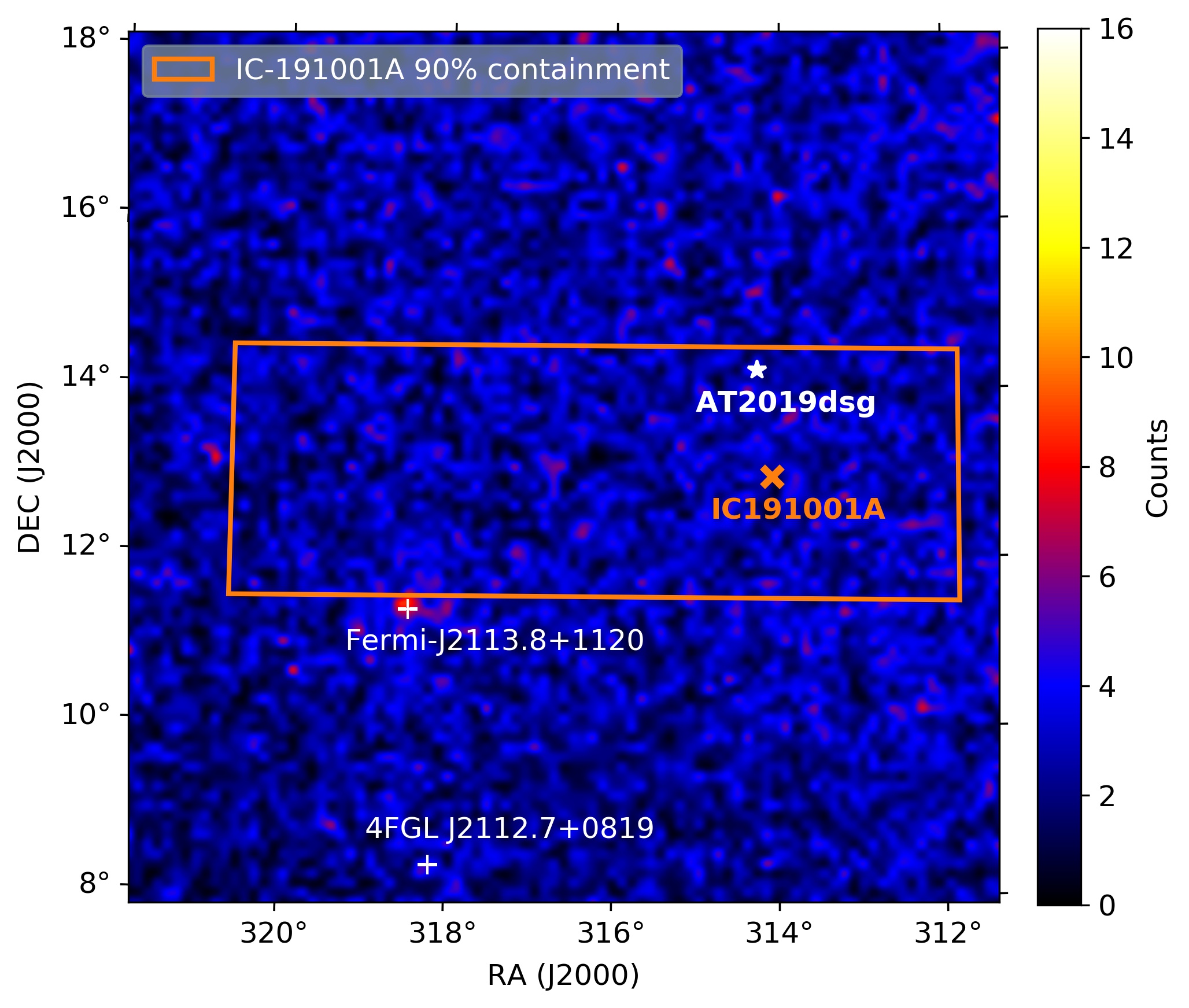}
    \caption{\textbf{LAT count map of the Region Of Interest (ROI)}. The map shows the integrated search period G3, showing the IC191001A 90$\%$ localisation region in orange. The position of AT2019dsg is marked by a white star. The neutrino best-fit position is marked with a orange `$\times$'. Two gamma-ray sources are significantly detected ($\geq$ 5 $\sigma$) in the ROI but outside the neutrino uncertainty region as marked with white crosses. There is no excess consistent with the position of AT2019dsg. %Gamma-ray sources significantly detected ($\geq$ 5 $\sigma$) are marked with white crosses. 
    }
    \label{fig:fermi}
\end{figure}
\newpage

\begin{figure}
    \centering
    \includegraphics[width=\textwidth]{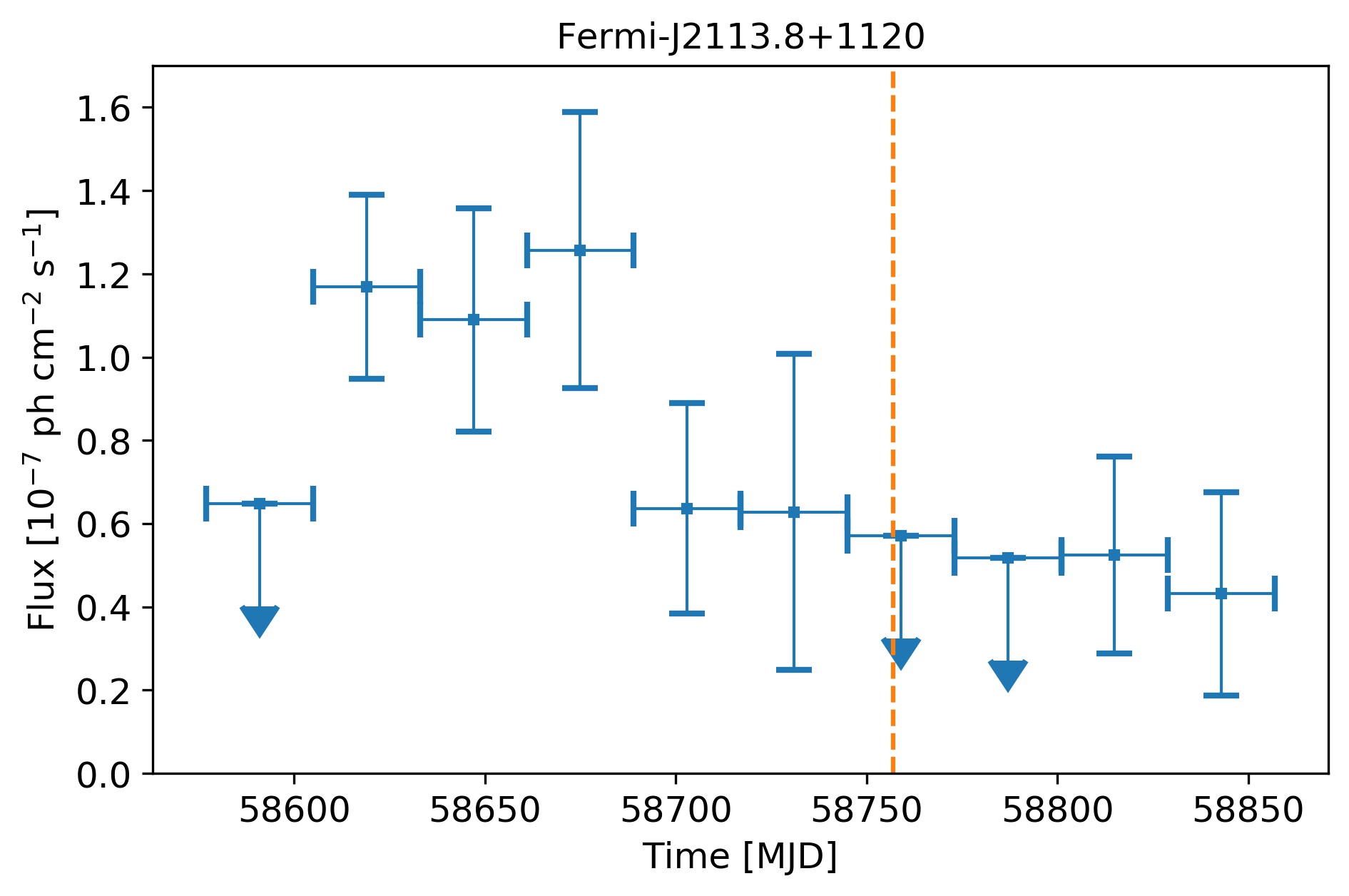}
    \caption{\textbf{LAT lightcurve for \textit{Fermi}-J2113.8+1120}.  The flux is derived in the 0.1-800 GeV energy range for the source during the time interval G3, with evenly spaced binning of 28 days. Vertical error bars represent 1$\sigma$ intervals, horizontal bars denote bin width. 2$\sigma$ upper limits are shown for bins with TS$\leq$9. The orange dashed vertical line marks the arrival time of IC-191001A. Since this source lies outside the reported 90\% error region (see Extended Data Figure 4), and given that the LAT lightcurve shows no obvious correlation with the neutrino arrival time, we conclude that it is unlikely to be associated with the neutrino.}
    \label{fig:fermi_lc}
\end{figure}
\clearpage

\begin{figure}
\centering
	\includegraphics[width=\textwidth]{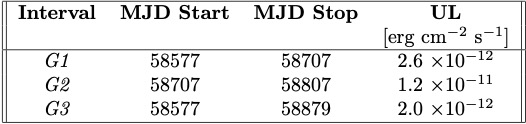}
	\caption{\textbf{Gamma-ray energy flux upper-limits for AT2019dsg.} The values are derived assuming a point-source with power-law index $\Gamma$=2.0 at the position of AT2019dsg, integrated over the analysis energy range 0.1-800 GeV.}
	\label{tab:lat_uls}
\end{figure}
\newpage

\begin{figure}
%\footnotesize
\centering
\includegraphics{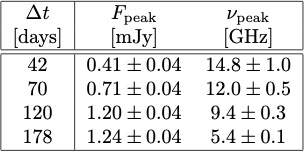}
\caption{\textbf{Peak frequency and peak flux density of the radio observations.} The time ($\Delta t$) is measured in the observer frame relative to MJD 58582.8, the date of discovery for AT2019dsg.}
\label{tab:syncobs}
\end{figure}

\begin{figure}
%\footnotesize
\includegraphics[width=\textwidth]{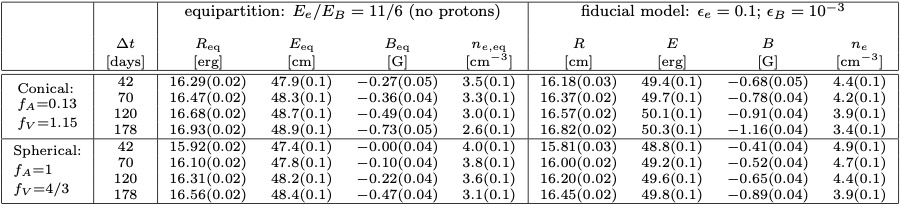}
\caption{\textbf{Properties of radio-emitting region.} These values are inferred from the synchrotron peak flux and peak frequency (see Extended Data Figure 7), where R is the region radius, E is the non-thermal energy, B is the magnetic field strength and $n_{e}$ is the density of non-thermal electrons. Except for $\Delta t$, all quantities are reported as log$_{10}$ with the uncertainty (68\% CL) listed in brackets.}
\label{tab:syncinf}
\end{figure}
\clearpage

\end{methods}


\begin{thebibliography}{10}
\providecommand{\url}[1]{\texttt{#1}}
\providecommand{\urlprefix}{URL }

\bibitem{2019TNSTR.615....1N}
{Nordin}, J. et~al.
\newblock {ZTF Transient Discovery Report for 2019-04-22}.
\newblock \emph{Transient Name Server Discovery Report} \textbf{2019-615} 1
  (2019).

\bibitem{2019ATel12752....1N}
{Nicholl}, M. et~al.
\newblock {ePESSTO+ classification of optical transients}.
\newblock \emph{The Astronomer's Telegram} \textbf{12752} 1 (2019).

\bibitem{2019ATel12798....1S}
{Sfaradi}, I. et~al.
\newblock {A possible radio detection of the TDE candidate AT2019DSG by
  AMI-LA}.
\newblock \emph{The Astronomer's Telegram} \textbf{12798} 1 (2019).

\bibitem{2019ATel12960....1P}
{Perez-Torres}, M. et~al.
\newblock {Unambiguous radio detection of the tidal disruption event AT2019dsg
  with e-MERLIN}.
\newblock \emph{The Astronomer's Telegram} \textbf{12960} 1 (2019).

\bibitem{2020arXiv200101409V}
{van Velzen}, S. et~al.
\newblock {Seventeen Tidal Disruption Events from the First Half of ZTF Survey
  Observations: Entering a New Era of Population Studies}.
\newblock \emph{arXiv e-prints} arXiv:2001.01409 (2020).

\bibitem{1984ARA&A..22..425H}
{Hillas}, A.M.
\newblock {The Origin of Ultra-High-Energy Cosmic Rays}.
\newblock \emph{\araa} \textbf{22} 425--444 (1984).

\bibitem{2017ApJ...838....3S}
{Senno}, N., {Murase}, K. \& {M{\'e}sz{\'a}ros}, P.
\newblock {High-energy Neutrino Flares from X-Ray Bright and Dark Tidal
  Disruption Events}.
\newblock \emph{\apj} \textbf{838} 3 (2017).

\bibitem{2017PhRvD..95l3001L}
{Lunardini}, C. \& {Winter}, W.
\newblock {High energy neutrinos from the tidal disruption of stars}.
\newblock \emph{\prd} \textbf{95} 123001 (2017).

\bibitem{2018Sci...361.1378I}
{IceCube Collaboration} et~al.
\newblock {Multimessenger observations of a flaring blazar coincident with
  high-energy neutrino IceCube-170922A}.
\newblock \emph{Science} \textbf{361} eaat1378 (2018).

\bibitem{2010ApJ...721..630H}
{H{\"u}mmer}, S., {R{\"u}ger}, M., {Spanier}, F. \& {Winter}, W.
\newblock {Simplified Models for Photohadronic Interactions in Cosmic
  Accelerators}.
\newblock \emph{\apj} \textbf{721} 630--652 (2010).

\bibitem{Waxman:1998yy}
{Waxman}, E. \& {Bahcall}, J.
\newblock {High energy neutrinos from astrophysical sources: An upper bound}.
\newblock \emph{\prd} \textbf{59} 023002 (1999).

\bibitem{2019ICRC...36.1016S}
{Stein}, R.
\newblock {Search for High-Energy Neutrinos from Populations of Optical
  Transients}.
\newblock In \emph{36th International Cosmic Ray Conference (ICRC2019)},
  volume~36 of \emph{International Cosmic Ray Conference}, 1016 (2019).

\bibitem{Sun:2015bda}
Sun, H., Zhang, B. \& Li, Z.
\newblock {Extragalactic High-energy Transients: Event Rate Densities and
  Luminosity Functions}.
\newblock \emph{Astrophys. J.} \textbf{812} 33 (2015).

\bibitem{flarestack}
Stein, R., Necker, J., Bradascio, F. \& Garrappa, S.
\newblock IceCubeOpenSource/flarestack: Titan V2.2.3 (2020).
\newblock \urlprefix\url{https://doi.org/10.5281/zenodo.4005800}.

\bibitem{ps_iceube_19}
{Pan-Starrs Collaboration} et~al.
\newblock {Search for transient optical counterparts to high-energy IceCube
  neutrinos with Pan-STARRS1}.
\newblock \emph{\aap} \textbf{626} A117 (2019).

\bibitem{blaufuss:gcn24378}
{Blaufuss}, E.
\newblock {IceCube-190503A - IceCube observation of a high-energy neutrino
  candidate event}.
\newblock \emph{GCN Circular} \textbf{24378} (2019).

\bibitem{2019ATel12730....1S}
{Stein}, R. et~al.
\newblock {Optical follow-up of IceCube-190503A with ZTF}.
\newblock \emph{The Astronomer's Telegram} \textbf{12730} 1 (2019).

\bibitem{blaufuss:gcn24910}
{Blaufuss}, E.
\newblock {IceCube-190629A - IceCube observation of a high-energy neutrino
  candidate event}.
\newblock \emph{GCN Circular} \textbf{24910} (2019).

\bibitem{2019ATel12879....1S}
{Stein}, R. et~al.
\newblock {Optical follow-up of IceCube-190619A with ZTF}.
\newblock \emph{The Astronomer's Telegram} \textbf{12879} 1 (2019).

\bibitem{stein:gcn25225}
{Stein}, R.
\newblock {IceCube-190730A - IceCube observation of a high-energy neutrino
  candidate event}.
\newblock \emph{GCN Circular} \textbf{25225} (2019).

\bibitem{2019ATel12974....1S}
{Stein}, R. et~al.
\newblock {Optical follow-up of IceCube-190730A with ZTF}.
\newblock \emph{The Astronomer's Telegram} \textbf{12974} 1 (2019).

\bibitem{blaufuss:gcn25806}
{Blaufuss}, E.
\newblock {IceCube-190922B - IceCube observation of a high-energy neutrino
  candidate event}.
\newblock \emph{GCN Circular} \textbf{25806} (2019).

\bibitem{2019ATel13125....1S}
{Stein}, R., {Franckowiak}, A., {Kowalski}, M. \& {Kasliwal}, M.
\newblock {A candidate supernova coincident with IceCube-190922B from ZTF}.
\newblock \emph{The Astronomer's Telegram} \textbf{13125} 1 (2019).

\bibitem{stein:gcn25824}
{Stein}, R., {Franckowiak}, A., {Kowalski}, M. \& {Kasliwal}, M.
\newblock {IceCube-190922B: Identification of a Candidate Supernova from the
  Zwicky Transient Facility}.
\newblock \emph{GCN Circular} \textbf{25824} (2019).

\bibitem{stein:gcn25913}
{Stein}, R.
\newblock {IceCube-191001A - IceCube observation of a high-energy neutrino
  candidate event}.
\newblock \emph{GCN Circular} \textbf{25913} (2019).

\bibitem{2019ATel13160....1S}
{Stein}, R., {Franckowiak}, A., {Necker}, J., {Gezari}, S. \& {Velzen}, S.v.
\newblock {Candidate Counterparts to IceCube-191001A with ZTF}.
\newblock \emph{The Astronomer's Telegram} \textbf{13160} 1 (2019).

\bibitem{stein:gcn25929}
{Stein}, R., {Franckowiak}, A., {Necker}, J. \& Suvi~{Gezari}, S.v.
\newblock {IceCube-191001A: Candidate Counterparts with the Zwicky Transient
  Facility}.
\newblock \emph{GCN Circular} \textbf{25929} (2019).

\bibitem{stein:gcn26655}
{Stein}, R.
\newblock {IceCube-200107A: IceCube observation of a high-energy neutrino
  candidate event}.
\newblock \emph{GCN Circular} \textbf{26655} (2020).

\bibitem{stein:gcn26667}
{Stein}, R. \& {Reusch}, S.
\newblock {IceCube-200107A: No candidates from the Zwicky Transient Facility}.
\newblock \emph{GCN Circular} \textbf{26667} (2020).

\bibitem{stein:gcn26696}
{Stein}, R.
\newblock {IceCube-200109A: IceCube observation of a high-energy neutrino
  candidate event}.
\newblock \emph{GCN Circular} \textbf{26696} (2020).

\bibitem{reusch:gcn26747}
{Reusch}, S. \& {Stein}, R.
\newblock {IceCube-200109A: Candidate Counterparts from the Zwicky Transient
  Facility}.
\newblock \emph{GCN Circular} \textbf{26747} (2020).

\bibitem{lagunas:gcn26802}
{Lagunas Gualda}, C.
\newblock {IceCube-200117A: IceCube observation of a high-energy neutrino
  candidate event}.
\newblock \emph{GCN Circular} \textbf{26802} (2020).

\bibitem{reusch:gcn26813}
{Reusch}, S. \& {Stein}, R.
\newblock {IceCube-200117A: Candidate Counterpart from the Zwicky Transient
  Facility}.
\newblock \emph{GCN Circular} \textbf{26813} (2020).

\bibitem{reusch:gcn26816}
{Reusch}, S. \& {Stein}, R.
\newblock {IceCube-200117A: One Additional Candidate Counterpart from the
  Zwicky Transient Facility}.
\newblock \emph{GCN Circular} \textbf{26816} (2020).

\bibitem{IC180423A}
{Kopper}, C.
\newblock {Retraction of IceCube GCN/AMON NOTICE 71165249{\_}130949}.
\newblock \emph{GCN Circular} \textbf{22669} (2018).

\bibitem{IC181031A}
{Blaufuss}, E.
\newblock {IceCube-181031A retraction}.
\newblock \emph{GCN Circular} \textbf{23398} (2018).

\bibitem{IC190205A}
{Blaufuss}, E.
\newblock {Retraction of IceCube GCN/AMON NOTICE 36142391{\_}132143}.
\newblock \emph{GCN Circular} \textbf{23876} (2019).

\bibitem{IC190529A}
{Blaufuss}, E.
\newblock {IceCube 41485283{\_}132628.amon retraction}.
\newblock \emph{GCN Circular} \textbf{24674} (2019).

\bibitem{IC180908A}
{Blaufuss}, E.
\newblock {IceCube-180908A - IceCube observation of a high-energy neutrino
  candidate event}.
\newblock \emph{GCN Circular} \textbf{23214} (2018).

\bibitem{IC181014A}
{Taboada}, I.
\newblock {IceCube-181014A - IceCube observation of a high-energy neutrino
  candidate event}.
\newblock \emph{GCN Circular} \textbf{23338} (2018).

\bibitem{IC190124A}
{Blaufuss}, E.
\newblock {IceCube-190124A - IceCube observation of a high-energy neutrino
  candidate event}.
\newblock \emph{GCN Circular} \textbf{23785} (2019).

\bibitem{IC190704A}
{Santander}, M.
\newblock {IceCube-190704A - IceCube observation of a high-energy neutrino
  candidate event}.
\newblock \emph{GCN Circular} \textbf{24981} (2019).

\bibitem{IC190712A}
{Blaufuss}, E.
\newblock {IceCube-190712A - IceCube observation of a high-energy neutrino
  candidate event}.
\newblock \emph{GCN Circular} \textbf{25057} (2019).

\bibitem{IC190819A}
{Santander}, M.
\newblock {IceCube-190819A - IceCube observation of a high-energy neutrino
  candidate event}.
\newblock \emph{GCN Circular} \textbf{25402} (2019).

\bibitem{IC191119A}
{Blaufuss}, E.
\newblock {IceCube-191119A - IceCube observation of a high-energy neutrino
  candidate event}.
\newblock \emph{GCN Circular} \textbf{26258} (2019).

\bibitem{IC200227A}
{Stein}, R.
\newblock {IceCube-200227A: IceCube observation of a high-energy neutrino
  candidate event}.
\newblock \emph{GCN Circular} \textbf{27235} (2020).

\bibitem{IC191215A}
{Stein}, R.
\newblock {IceCube-191215A - IceCube observation of a high-energy neutrino
  candidate event}.
\newblock \emph{GCN Circular} \textbf{26435} (2019).

\bibitem{IC190331A}
{Kopper}, C.
\newblock {IceCube-190331A - IceCube observation of a high-energy neutrino
  candidate event}.
\newblock \emph{GCN Circular} \textbf{24028} (2019).

\bibitem{IC190504A}
{Kopper}, C.
\newblock {IceCube-190504A - IceCube observation of a high-energy neutrino
  candidate event}.
\newblock \emph{GCN Circular} \textbf{24392} (2019).

\bibitem{IC190221A}
{Taboada}, I.
\newblock {IceCube-190921A - IceCube observation of a high-energy neutrino
  candidate event}.
\newblock \emph{GCN Circular} \textbf{23918} (2019).

\bibitem{IC190629A}
{Blaufuss}, E.
\newblock {IceCube-190629A - IceCube observation of a high-energy neutrino
  candidate event}.
\newblock \emph{GCN Circular} \textbf{24910} (2019).

\bibitem{IC190922A}
{Stein}, R.
\newblock {IceCube-190922A - IceCube observation of a high-energy neutrino
  candidate event}.
\newblock \emph{GCN Circular} \textbf{25802} (2019).

\bibitem{IC191122A}
{Blaufuss}, E.
\newblock {IceCube-191122A - IceCube observation of a high-energy neutrino
  candidate event}.
\newblock \emph{GCN Circular} \textbf{26276} (2019).

\bibitem{IC191204A}
{Stein}, R.
\newblock {IceCube-191204A - IceCube observation of a high-energy neutrino
  candidate event}.
\newblock \emph{GCN Circular} \textbf{26341} (2019).

\bibitem{IC191231A}
{Santander}, M.
\newblock {IceCube-191231A: IceCube observation of a high-energy neutrino
  candidate event}.
\newblock \emph{GCN Circular} \textbf{26620} (2019).

\bibitem{IC200120A}
{Lagunas Gualda}, C.
\newblock {IceCube-200120A: IceCube observation of a high-energy neutrino
  candidate event}.
\newblock \emph{GCN Circular} \textbf{26832} (2020).

\bibitem{IC200120A_2}
{Blaufuss}, E.
\newblock {IceCube-200120A: Event likely due to background}.
\newblock \emph{GCN Circular} \textbf{26874} (2020).

\bibitem{IC181023A}
{Blaufuss}, E.
\newblock {IceCube-181023A - IceCube observation of a high-energy neutrino
  candidate event}.
\newblock \emph{GCN Circular} \textbf{23375} (2018).

\end{thebibliography}


\begin{thebibliography}{10}
	
\providecommand{\url}[1]{\texttt{#1}}
\providecommand{\urlprefix}{URL }

\bibitem{icecube_detector}
{Aartsen}, M.G. et~al.
\newblock {The IceCube Neutrino Observatory: instrumentation and online
	systems}.
\newblock \emph{Journal of Instrumentation} \textbf{12} P03012 (2017).

\bibitem{stein:gcn25913}
{Stein}, R.
\newblock {IceCube-191001A - IceCube observation of a high-energy neutrino
	candidate event}.
\newblock \emph{GCN Circular} \textbf{25913} (2019).

\bibitem{2019PASP..131a8002B}
{Bellm}, E.C. et~al.
\newblock {The Zwicky Transient Facility: System Overview, Performance, and
	First Results}.
\newblock \emph{\pasp} \textbf{131} 018002 (2019).

\bibitem{2007APh....27..533K}
{Kowalski}, M. \& {Mohr}, A.
\newblock {Detecting neutrino transients with optical follow-up observations}.
\newblock \emph{Astroparticle Physics} \textbf{27} 533--538 (2007).

\bibitem{2009ApJ...693..329F}
{Farrar}, G.R. \& {Gruzinov}, A.
\newblock {Giant AGN Flares and Cosmic Ray Bursts}.
\newblock \emph{\apj} \textbf{693} 329--332 (2009).

\bibitem{2017MNRAS.469.1354D}
{Dai}, L. \& {Fang}, K.
\newblock {Can tidal disruption events produce the IceCube neutrinos?}
\newblock \emph{\mnras} \textbf{469} 1354--1359 (2017).

\bibitem{2019ApJ...886..114H}
{Hayasaki}, K. \& {Yamazaki}, R.
\newblock {Neutrino Emissions from Tidal Disruption Remnants}.
\newblock \emph{\apj} \textbf{886} 114 (2019).

\bibitem{2014arXiv1411.0704F}
{Farrar}, G.R. \& {Piran}, T.
\newblock {Tidal disruption jets as the source of Ultra-High Energy Cosmic
	Rays}.
\newblock \emph{arXiv e-prints} arXiv:1411.0704 (2014).

\bibitem{2017ApJ...838....3S}
{Senno}, N., {Murase}, K. \& {M{\'e}sz{\'a}ros}, P.
\newblock {High-energy Neutrino Flares from X-Ray Bright and Dark Tidal
	Disruption Events}.
\newblock \emph{\apj} \textbf{838} 3 (2017).

\bibitem{2016PhRvD..93h3005W}
{Wang}, X.Y. \& {Liu}, R.Y.
\newblock {Tidal disruption jets of supermassive black holes as hidden sources
	of cosmic rays: Explaining the IceCube TeV-PeV neutrinos}.
\newblock \emph{\prd} \textbf{93} 083005 (2016).

\bibitem{2017PhRvD..95l3001L}
{Lunardini}, C. \& {Winter}, W.
\newblock {High energy neutrinos from the tidal disruption of stars}.
\newblock \emph{\prd} \textbf{95} 123001 (2017).

\bibitem{2019ATel13160....1S}
{Stein}, R., {Franckowiak}, A., {Necker}, J., {Gezari}, S. \& {Velzen}, S.v.
\newblock {Candidate Counterparts to IceCube-191001A with ZTF}.
\newblock \emph{The Astronomer's Telegram} \textbf{13160} 1 (2019).

\bibitem{2019PASP..131g8001G}
{Graham}, M.J. et~al.
\newblock {The Zwicky Transient Facility: Science Objectives}.
\newblock \emph{\pasp} \textbf{131} 078001 (2019).

\bibitem{2019TNSTR.615....1N}
{Nordin}, J. et~al.
\newblock {ZTF Transient Discovery Report for 2019-04-22}.
\newblock \emph{Transient Name Server Discovery Report} \textbf{2019-615} 1
(2019).

\bibitem{2019ATel12752....1N}
{Nicholl}, M. et~al.
\newblock {ePESSTO+ classification of optical transients}.
\newblock \emph{The Astronomer's Telegram} \textbf{12752} 1 (2019).

\bibitem{2020arXiv200101409V}
{van Velzen}, S. et~al.
\newblock {Seventeen Tidal Disruption Events from the First Half of ZTF Survey
	Observations: Entering a New Era of Population Studies}.
\newblock \emph{arXiv e-prints} arXiv:2001.01409 (2020).

\bibitem{2019ApJ...878...82V}
{van Velzen}, S. et~al.
\newblock {Late-time UV Observations of Tidal Disruption Flares Reveal
	Unobscured, Compact Accretion Disks}.
\newblock \emph{\apj} \textbf{878} 82 (2019).

\bibitem{2020MNRAS.492.5655M}
{Mummery}, A. \& {Balbus}, S.A.
\newblock {The spectral evolution of disc dominated tidal disruption events}.
\newblock \emph{\mnras} \textbf{492} 5655--5674 (2020).

\bibitem{2013ApJ...764..184M}
{McConnell}, N.J. \& {Ma}, C.P.
\newblock {Revisiting the Scaling Relations of Black Hole Masses and Host
	Galaxy Properties}.
\newblock \emph{\apj} \textbf{764} 184 (2013).

\bibitem{2017ApJ...838..149A}
{Auchettl}, K., {Guillochon}, J. \& {Ramirez-Ruiz}, E.
\newblock {New Physical Insights about Tidal Disruption Events from a
	Comprehensive Observational Inventory at X-Ray Wavelengths}.
\newblock \emph{\apj} \textbf{838} 149 (2017).

\bibitem{2019MNRAS.487.4136W}
{Wevers}, T. et~al.
\newblock {Black hole masses of tidal disruption event host galaxies II}.
\newblock \emph{\mnras} \textbf{487} 4136--4152 (2019).

\bibitem{2019ApJ...872..198V}
{van Velzen}, S. et~al.
\newblock {The First Tidal Disruption Flare in ZTF: From Photometric Selection
	to Multi-wavelength Characterization}.
\newblock \emph{\apj} \textbf{872} 198 (2019).

\bibitem{2012A&A...538A..81M}
{Morlino}, G. \& {Caprioli}, D.
\newblock {Strong evidence for hadron acceleration in Tycho's supernova
	remnant}.
\newblock \emph{\aap} \textbf{538} A81 (2012).

\bibitem{2018ApJ...854...86E}
{Eftekhari}, T., {Berger}, E., {Zauderer}, B.A., {Margutti}, R. \& {Alexander},
K.D.
\newblock {Radio Monitoring of the Tidal Disruption Event Swift
	J164449.3+573451. III. Late-time Jet Energetics and a Deviation from
	Equipartition}.
\newblock \emph{\apj} \textbf{854} 86 (2018).

\bibitem{2013MNRAS.436.1258H}
{Horesh}, A. et~al.
\newblock {An early and comprehensive millimetre and centimetre wave and X-ray
	study of SN 2011dh: a non-equipartition blast wave expanding into a massive
	stellar wind}.
\newblock \emph{\mnras} \textbf{436} 1258--1267 (2013).

\bibitem{2013ApJ...772...78B}
{Barniol Duran}, R., {Nakar}, E. \& {Piran}, T.
\newblock {Radius Constraints and Minimal Equipartition Energy of
	Relativistically Moving Synchrotron Sources}.
\newblock \emph{\apj} \textbf{772} 78 (2013).

\bibitem{2003PASA...20...69P}
{Polatidis}, A.G. \& {Conway}, J.E.
\newblock {Proper Motions in Compact Symmetric Objects}.
\newblock \emph{\pasa} \textbf{20} 69--74 (2003).

\bibitem{2016ApJ...819L..25A}
{Alexander}, K.D., {Berger}, E., {Guillochon}, J., {Zauderer}, B.A. \&
{Williams}, P.K.G.
\newblock {Discovery of an Outflow from Radio Observations of the Tidal
	Disruption Event ASASSN-14li}.
\newblock \emph{\apjl} \textbf{819} L25 (2016).

\bibitem{2016ApJ...827..127K}
{Krolik}, J., {Piran}, T., {Svirski}, G. \& {Cheng}, R.M.
\newblock {ASASSN-14li: A Model Tidal Disruption Event}.
\newblock \emph{\apj} \textbf{827} 127 (2016).

\bibitem{2018ApJ...856....1P}
{Pasham}, D.R. \& {van Velzen}, S.
\newblock {Discovery of a Time Lag between the Soft X-Ray and Radio Emission of
	the Tidal Disruption Flare ASASSN-14li: Evidence for Linear Disk-Jet
	Coupling}.
\newblock \emph{\apj} \textbf{856} 1 (2018).

\bibitem{2019A&A...622L...9S}
{Strotjohann}, N.L., {Kowalski}, M. \& {Franckowiak}, A.
\newblock {Eddington bias for cosmic neutrino sources}.
\newblock \emph{\aap} \textbf{622} L9 (2019).

\bibitem{1984ARA&A..22..425H}
{Hillas}, A.M.
\newblock {The Origin of Ultra-High-Energy Cosmic Rays}.
\newblock \emph{\araa} \textbf{22} 425--444 (1984).

\bibitem{2018Sci...361.1378I}
{IceCube Collaboration} et~al.
\newblock {Multimessenger observations of a flaring blazar coincident with
	high-energy neutrino IceCube-170922A}.
\newblock \emph{Science} \textbf{361} eaat1378 (2018).

\bibitem{2019ICRC...36.1021B}
{Blaufuss}, E., {Kintscher}, T., {Lu}, L. \& {Tung}, C.F.
\newblock {The Next Generation of IceCube Real-time Neutrino Alerts}.
\newblock In \emph{36th International Cosmic Ray Conference (ICRC2019)},
volume~36 of \emph{International Cosmic Ray Conference}, 1021 (2019).

\bibitem{2016PhRvL.116g1101M}
{Murase}, K., {Guetta}, D. \& {Ahlers}, M.
\newblock {Hidden Cosmic-Ray Accelerators as an Origin of TeV-PeV Cosmic
	Neutrinos}.
\newblock \emph{\prl} \textbf{116} 071101 (2016).

\bibitem{2019ICRC...36.1016S}
{Stein}, R.
\newblock {Search for High-Energy Neutrinos from Populations of Optical
	Transients}.
\newblock In \emph{36th International Cosmic Ray Conference (ICRC2019)},
volume~36 of \emph{International Cosmic Ray Conference}, 1016 (2019).
	
\setcounter{firstbib}{\value{enumiv}}
	
\end{thebibliography}

\noindent

\begin{thebibliography}{100}
	\setcounter{enumiv}{\value{firstbib}} 
	
	\providecommand{\url}[1]{\texttt{#1}}
	\providecommand{\urlprefix}{URL }
 
	 \bibitem{2019PASP..131d8001C}
	 {Coughlin}, M.W. et~al.
	 \newblock {2900 Square Degree Search for the Optical Counterpart of Short
	 	Gamma-Ray Burst GRB 180523B with the Zwicky Transient Facility}.
	 \newblock \emph{\pasp} \textbf{131} 048001 (2019).
	 
	 \bibitem{stein:gcn26655}
	 {Stein}, R.
	 \newblock {IceCube-200107A: IceCube observation of a high-energy neutrino
	 	candidate event}.
	 \newblock \emph{GCN Circular} \textbf{26655} (2020).
	 
	 \bibitem{2019PASP..131a8003M}
	 {Masci}, F.J. et~al.
	 \newblock {The Zwicky Transient Facility: Data Processing, Products, and
	 	Archive}.
	 \newblock \emph{\pasp} \textbf{131} 018003 (2019).
	 
	 \bibitem{2019PASP..131a8001P}
	 {Patterson}, M.T. et~al.
	 \newblock {The Zwicky Transient Facility Alert Distribution System}.
	 \newblock \emph{\pasp} \textbf{131} 018001 (2019).
	 
	 \bibitem{ampel_followup_pipeline}
	 Stein, R. \& Reusch, S.
	 \newblock robertdstein/ampel\_followup\_pipeline: V1.1 Release (2020).
	 \newblock \urlprefix\url{https://doi.org/10.5281/zenodo.4048336}.
	 
	 \bibitem{2019A&A...631A.147N}
	 {Nordin}, J. et~al.
	 \newblock {Transient processing and analysis using AMPEL: alert management,
	 	photometry, and evaluation of light curves}.
	 \newblock \emph{\aap} \textbf{631} A147 (2019).
	 
	 \bibitem{2019PASP..131c8002M}
	 {Mahabal}, A. et~al.
	 \newblock {Machine Learning for the Zwicky Transient Facility}.
	 \newblock \emph{\pasp} \textbf{131} 038002 (2019).
	 
	 \bibitem{2018PASP..130g5002S}
	 {Soumagnac}, M.T. \& {Ofek}, E.O.
	 \newblock {catsHTM: A Tool for Fast Accessing and Cross-matching Large
	 	Astronomical Catalogs}.
	 \newblock \emph{\pasp} \textbf{130} 075002 (2018).
	 
	 \bibitem{2018A&A...616A...1G}
	 {Gaia Collaboration} et~al.
	 \newblock {Gaia Data Release 2. Summary of the contents and survey properties}.
	 \newblock \emph{\aap} \textbf{616} A1 (2018).
	 
	 \bibitem{2018PASP..130l8001T}
	 {Tachibana}, Y. \& {Miller}, A.A.
	 \newblock {A Morphological Classification Model to Identify Unresolved
	 	PanSTARRS1 Sources: Application in the ZTF Real-time Pipeline}.
	 \newblock \emph{\pasp} \textbf{130} 128001 (2018).
	 
	 \bibitem{2016arXiv161205560C}
	 {Chambers}, K.C. et~al.
	 \newblock {The Pan-STARRS1 Surveys}.
	 \newblock \emph{arXiv e-prints} arXiv:1612.05560 (2016).
	 
	 \bibitem{2010AJ....140.1868W}
	 {Wright}, E.L. et~al.
	 \newblock {The Wide-field Infrared Survey Explorer (WISE): Mission Description
	 	and Initial On-orbit Performance}.
	 \newblock \emph{\aj} \textbf{140} 1868--1881 (2010).
	 
	 \bibitem{2020PhRvL.124e1103A}
	 {Aartsen}, M.G. et~al.
	 \newblock {Time-Integrated Neutrino Source Searches with 10 Years of IceCube
	 	Data}.
	 \newblock \emph{\prl} \textbf{124} 051103 (2020).
	 
	 \bibitem{2004SPIE.5489..679S}
	 {Steele}, I.A. et~al.
	 \newblock \emph{The Liverpool Telescope: performance and first results}, volume
	 5489 of \emph{Society of Photo-Optical Instrumentation Engineers (SPIE)
	 	Conference Series}, 679--692 (2004).
	 
	 \bibitem{Blagorodnova18}
	 {Blagorodnova}, N. et~al.
	 \newblock {The SED Machine: A Robotic Spectrograph for Fast Transient
	 	Classification}.
	 \newblock \emph{\pasp} \textbf{130} 035003 (2018).
	 
	 \bibitem{Rigault19}
	 {Rigault}, M. et~al.
	 \newblock {Fully automated integral field spectrograph pipeline for the
	 	SEDMachine: pysedm}.
	 \newblock \emph{\aap} \textbf{627} A115 (2019).
	 
	 \bibitem{fst+2016}
	 {Fremling}, C. et~al.
	 \newblock {PTF12os and iPTF13bvn. Two stripped-envelope supernovae from
	 	low-mass progenitors in NGC 5806}.
	 \newblock \emph{\aap} \textbf{593} A68 (2016).
	 
	 \bibitem{2018ApJ...852...72V}
	 {van Velzen}, S.
	 \newblock {On the Mass and Luminosity Functions of Tidal Disruption Flares:
	 	Rate Suppression due to Black Hole Event Horizons}.
	 \newblock \emph{\apj} \textbf{852} 72 (2018).
	 
	 \bibitem{2005SSRv..120...95R}
	 {Roming}, P.W.A. et~al.
	 \newblock {The Swift Ultra-Violet/Optical Telescope}.
	 \newblock \emph{\ssr} \textbf{120} 95--142 (2005).
	 
	 \bibitem{2004ApJ...611.1005G}
	 {Gehrels}, N. et~al.
	 \newblock {The Swift Gamma-Ray Burst Mission}.
	 \newblock \emph{\apj} \textbf{611} 1005--1020 (2004).
	 
	 \bibitem{2016ApJ...829...19V}
	 {van Velzen}, S., {Mendez}, A.J., {Krolik}, J.H. \& {Gorjian}, V.
	 \newblock {Discovery of Transient Infrared Emission from Dust Heated by Stellar
	 	Tidal Disruption Flares}.
	 \newblock \emph{\apj} \textbf{829} 19 (2016).
	 
	 \bibitem{2016MNRAS.458..575L}
	 {Lu}, W., {Kumar}, P. \& {Evans}, N.J.
	 \newblock {Infrared emission from tidal disruption events - probing the
	 	pc-scale dust content around galactic nuclei}.
	 \newblock \emph{\mnras} \textbf{458} 575--581 (2016).
	 
	 \bibitem{Miller93}
	 {Miller}, J.S. \& {Stone}, R.P.S.
	 \newblock \emph{{Lick Obs. Tech. Rep. 66}}.
	 \newblock Santa Cruz: Lick Obs. (1993).
	 
	 \bibitem{Oke95}
	 {Oke}, J.B. et~al.
	 \newblock {The Keck Low-Resolution Imaging Spectrometer}.
	 \newblock \emph{\pasp} \textbf{107} 375--+ (1995).
	 
	 \bibitem{2005MNRAS.359..765G}
	 {Garcia-Rissmann}, A. et~al.
	 \newblock {An atlas of calcium triplet spectra of active galaxies}.
	 \newblock \emph{\mnras} \textbf{359} 765--780 (2005).
	 
	 \bibitem{2005SSRv..120..165B}
	 {Burrows}, D.N. et~al.
	 \newblock {The Swift X-Ray Telescope}.
	 \newblock \emph{\ssr} \textbf{120} 165--195 (2005).
	 
	 \bibitem{2001A&A...365L...1J}
	 {Jansen}, F. et~al.
	 \newblock {XMM-Newton observatory. I. The spacecraft and operations}.
	 \newblock \emph{\aap} \textbf{365} L1--L6 (2001).
	 
	 \bibitem{HI4PI2016}
	 {HI4PI Collaboration} et~al.
	 \newblock {HI4PI: A full-sky H I survey based on EBHIS and GASS}.
	 \newblock \emph{\aap} \textbf{594} A116 (2016).
	 
	 \bibitem{1996ASPC..101...17A}
	 {Arnaud}, K.A.
	 \newblock \emph{{XSPEC: The First Ten Years}}, volume 101 of \emph{Astronomical
	 	Society of the Pacific Conference Series}, 17 (1996).
	 
	 \bibitem{2008MNRAS.391.1545Z}
	 {Zwart}, J.T.L. et~al.
	 \newblock {The Arcminute Microkelvin Imager}.
	 \newblock \emph{\mnras} \textbf{391} 1545--1558 (2008).
	 
	 \bibitem{2018MNRAS.475.5677H}
	 {Hickish}, J. et~al.
	 \newblock {A digital correlator upgrade for the Arcminute MicroKelvin Imager}.
	 \newblock \emph{\mnras} \textbf{475} 5677--5687 (2018).
	 
	 \bibitem{2015MNRAS.453.1396P}
	 {Perrott}, Y.C. et~al.
	 \newblock {AMI Galactic Plane Survey at 16 GHz - II. Full data release with
	 	extended coverage and improved processing}.
	 \newblock \emph{\mnras} \textbf{453} 1396--1403 (2015).
	 
	 \bibitem{McMullin2007}
	 {McMullin}, J.P., {Waters}, B., {Schiebel}, D., {Young}, W. \& {Golap}, K.
	 \newblock {CASA Architecture and Applications}.
	 \newblock In R.A. {Shaw}, F.~{Hill} \& D.J. {Bell}, editors, \emph{Astronomical
	 	Data Analysis Software and Systems XVI}, volume 376 of \emph{Astronomical
	 	Society of the Pacific Conference Series}, 127 (2007).
	 
	 \bibitem{2009ApJ...697.1071A}
	 {Atwood}, W.B., {Abdo}, A.A., {Ackermann}, M. et~al.
	 \newblock {The Large Area Telescope on the Fermi Gamma-Ray Space Telescope
	 	Mission}.
	 \newblock \emph{\apj} \textbf{697} 1071--1102 (2009).
	 
	 \bibitem{2017ICRC...35..824W}
	 {Wood}, M. et~al.
	 \newblock {Fermipy: An open-source Python package for analysis of Fermi-LAT
	 	Data}.
	 \newblock In \emph{35th International Cosmic Ray Conference (ICRC2017)}, volume
	 301 of \emph{International Cosmic Ray Conference}, 824 (2017).
	 
	 \bibitem{garrappa_buson:gcn25932}
	 {Garrappa}, S. \& {Buson}, S.
	 \newblock {Fermi-LAT Gamma-ray Observations of IceCube-191001A}.
	 \newblock \emph{GCN Circular} \textbf{25932} (2019).
	 
	 \bibitem{2019arXiv190210045T}
	 {The Fermi-LAT collaboration}.
	 \newblock {Fermi Large Area Telescope Fourth Source Catalog}.
	 \newblock \emph{arXiv e-prints} arXiv:1902.10045 (2019).
	 
	 \bibitem{2013ApJ...767...14P}
	 {Pursimo}, T. et~al.
	 \newblock {The Micro-Arcsecond Scintillation-Induced Variability (MASIV)
	 	Survey. III. Optical Identifications and New Redshifts}.
	 \newblock \emph{\apj} \textbf{767} 14 (2013).
	 
	 \bibitem{2019GCN.25932....1G}
	 {Garrappa}, S., {Buson}, S. \& {Fermi-LAT Collaboration}.
	 \newblock {Fermi-LAT Gamma-ray Observations of IceCube-191001A}.
	 \newblock \emph{GCN Circular} \textbf{25932} 1 (2019).
	 
	 \bibitem{2015ApJ...802..133D}
	 {Diltz}, C., {B{\"o}ttcher}, M. \& {Fossati}, G.
	 \newblock {Time Dependent Hadronic Modeling of Flat Spectrum Radio Quasars}.
	 \newblock \emph{\apj} \textbf{802} 133 (2015).
	 
	 \bibitem{2019NatAs...3...88G}
	 {Gao}, S., {Fedynitch}, A., {Winter}, W. \& {Pohl}, M.
	 \newblock {Modelling the coincident observation of a high-energy neutrino and a
	 	bright blazar flare}.
	 \newblock \emph{Nature Astronomy} \textbf{3} 88--92 (2019).
	 
	 \bibitem{ayala:gcn25936}
	 {Ayala}, H.
	 \newblock {IceCube-191001A: HAWC follow-up}.
	 \newblock \emph{GCN Circular} \textbf{25936} (2019).
	 
	 \bibitem{2016Sci...351...62V}
	 {van Velzen}, S. et~al.
	 \newblock {A radio jet from the optical and x-ray bright stellar tidal
	 	disruption flare ASASSN-14li}.
	 \newblock \emph{Science} \textbf{351} 62--65 (2016).
	 
	 \bibitem{Foreman-Mackey13}
	 {Foreman-Mackey}, D., {Hogg}, D.W., {Lang}, D. \& {Goodman}, J.
	 \newblock {emcee: The MCMC Hammer}.
	 \newblock \emph{\pasp} \textbf{125} 306 (2013).
	 
	 \bibitem{Guillochon18}
	 {Guillochon}, J. et~al.
	 \newblock {MOSFiT: Modular Open Source Fitter for Transients}.
	 \newblock \emph{\apjs} \textbf{236} 6 (2018).
	 
	 \bibitem{Granot14}
	 {Granot}, J. \& {van der Horst}, A.J.
	 \newblock {Gamma-Ray Burst Jets and their Radio Observations}.
	 \newblock \emph{\pasa} \textbf{31} e008 (2014).
	 
	 \bibitem{Fong15}
	 {Fong}, W., {Berger}, E., {Margutti}, R. \& {Zauderer}, B.A.
	 \newblock {A Decade of Short-duration Gamma-Ray Burst Broadband Afterglows:
	 	Energetics, Circumburst Densities, and Jet Opening Angles}.
	 \newblock \emph{\apj} \textbf{815} 102 (2015).

\end{thebibliography}
\end{document}

\maketitle
\newline

\begin{affiliations}
 \item {Deutsches Elektronen Synchrotron DESY, Platanenallee 6, 15738 Zeuthen, Germany}
 \item {Institut f{\"u}r Physik, Humboldt-Universit{\"a}t zu Berlin, D-12489 Berlin, Germany}
\item Center for Cosmology and Particle Physics, New York University, NY 10003, USA
\item Department of Astronomy, University of Maryland, College Park, MD 20742, USA
\item Leiden Observatory, Leiden University, PO Box 9513, 2300 RA Leiden, The Netherlands
\item {Columbia Astrophysics Laboratory, Columbia University in the City of New York, 550 W 120th St., New York, NY 10027, USA}
 \item Joint Space-Science Institute, University of Maryland, College Park, MD 20742, USA
 \item International Centre for Radio Astronomy Research -- Curtin University, Perth, Western Australia, Australia
 % Itai Sfaradi and Assaf Horesh
 \item {Racah Institute of Physics, The Hebrew University of Jerusalem, Jerusalem 91904, Israel} % 
 % M. Bietenholz, A. de Witt
 \item{Hartebeesthoek Radio Astronomy Observatory, SARAO, PO Box 443, Krugersdorp, 1740, South Africa }
 \item{Department of Physics and Astronomy, York University, Toronto, M3J~1P3, Ontario, Canada}
% Rob Fender, P. Wouldt
\item Astrophysics, Department of Physics, University of Oxford, Keble Road, OX1 3RH, UK
\item Department of Astronomy, University of Cape Town, Private Bag X3, Rondebosch, 7701, South
Africa
%  I. Andreoni, Y. Yao, M. Graham
\item {Division of Physics, Mathematics, and Astronomy, California Institute of Technology, 1200 East California Blvd, MC 249-17, Pasadena, CA 91125, USA}
% J. Belecki, R. Burrus, M. Feeney, H. Rodriguez
\item{Caltech Optical Observatories, California Institute of Technology, Pasadena, CA  91125, USA}
% E. Bellm + V. Zach Golkhou
\item{DIRAC Institute, Department of Astronomy, University of Washington, 3910 15th Avenue NE, Seattle, WA 98195, USA}
% M. Boettcher
\item{Centre for Space Research, North-West University, Potchefstroom, 2531,
South Africa}
% S. B. Cenko, L. P. Singer, 
\item{Astroparticle Physics Laboratory, NASA Goddard Space Flight Center, Mail Code 661, Greenbelt, MD 20771, USA}
%  M. Coughlin:
\item{School of Physics and Astronomy, University of Minnesota,
Minneapolis, Minnesota 55455, USA}
% Ryan Foley, Tiara Hung, 
\item{Department of Astronomy and Astrophysics, University of California, Santa Cruz, California, 95064, USA}
% Gal-Yam, Soumagnac
\item{Department of Particle Physics and Astrophysics, Weizmann Institute of Science, 234 Herzl St, Rehovot 76100, Israel}
% V. Zach Golkhou
\item{The eScience Institute, University of Washington, Seattle, WA 98195, USA}
% A. Goobar
\item{The Oskar Klein Centre, Department of Physics,  Stockholm University,
AlbaNova, SE-106 91 Stockholm, Sweden}
% Helou, Laher, Masci, Rusholme, Shupe
\item{IPAC, California Institute of Technology, 1200 E. California Blvd, Pasadena, CA 91125, USA}
% A. K. H. Kong
\item{Institute of Astronomy, National Tsing Hua University, No. 101 Section 2 Kuang-Fu Road, Hsinchu 30013, Taiwan}
% T. Kupfer
\item{Kavli Institute for Theoretical Physics, University of California, Santa Barbara, CA 93106, USA}
% A. Hamabal (+ caltech)
\item{Center for Data Driven Discovery, California Institute of Technology, Pasadena, CA 91125, USA}
% D. Perley + K. Taggart
\item{Astrophysics Research Institute, Liverpool John Moores University, 146 Brownlow Hill, Liverpool L3 5RF, UK
}
% Mickael Rigault
\item{Universit\'e Clermont Auvergne, CNRS/IN2P3, Laboratoire de Physique de Clermont, F-63000 Clermont-Ferrand, France}
% J. Sollerman
\item{The Oskar Klein Centre, Department of Astronomy,  Stockholm University,
AlbaNova, SE-106 91 Stockholm, Sweden}
% M. Soumagnac
\item{Lawrence Berkeley National Laboratory, 1 Cyclotron Road, Berkeley, CA 94720, USA}
% D. Stern
\item{Jet Propulsion Laboratory, California Institute of Technology, 4800 Oak Grove Drive, Pasadena, CA 91109, USA}
\end{affiliations}

\renewcommand{\thefigure}{S\arabic{figure}}
\renewcommand{\thetable}{S\arabic{table}}
\setcounter{figure}{0}

\section{Discovery and Classification History of AT2019dsg}
AT2019dsg was discovered by ZTF on 2019 April 9, initially named ZTF19aapreis, and reported on 2019 April 22 as a likely extragalactic transient\cite{2019TNSTR.615....1N}. AT2019dsg was subsequently classified as a TDE on 2019 May 13 by ePESSTO+\cite{2019ATel12752....1N}. Radio emission was tentatively reported on 2019 May 23 by AMI-LA\cite{2019ATel12798....1S}, and confirmed on 2019 July 26 by e-MERLIN\cite{2019ATel12960....1P}. In addition to observations as part of a systematic ZTF search for TDEs\cite{2020arXiv200101409V}, the association with IC191001A prompted additional follow-up. 

\section{Neutrino Production in AT2019dsg}
\label{sec:max_energy}
The Hillas criterion\cite{1984ARA&A..22..425H} for a system of magnetic field strength $B$ and physical radius $R$ can be expressed as\cite{1984ARA&A..22..425H}:

\begin{equation}
\frac{E_{\rm max}}{\rm PeV} \approx
1600 \times \frac{B}{\rm Gauss} \times \frac{R}{10^{16} \, \rm cm} \times
\beta Z
\end{equation}
where $Z$ is the particle charge, $\beta \sim 0.2$ is the outflow velocity in units of c and $E$ is the maximum charged-particle energy. In order for particle acceleration to occur, the timescale required for particle acceleration must be shorter than the associated particle cooling timescale. Previous work has found this condition can be satisfied in TDEs for relevant energies\cite{2017ApJ...838....3S, 2017PhRvD..95l3001L}, although a detailed calculation is beyond the scope of this work.

%\subsection{Target Density}
These accelerated protons then need sufficient target density. For a photon target, with p$\gamma$ pion production via the $\Delta$ resonance, we expect that neutrino production will occur above a threshold:

\begin{equation}
E_{\gamma}E_{p} \sim \Gamma ^{2} 0.16 \, \rm{GeV}^{2}
\end{equation} With this constraint, we can derive the necessary photon energies required for a target to produce IC191001A. Taking the reconstructed neutrino energy of $\sim$0.2 PeV directly, we find a threshold photon target of $E_{\gamma} > $40 eV. However, these reconstructed neutrino energies typically have upper bounds an order of magnitude or more above the central estimate\cite{2018Sci...361.1378I}, so the true neutrino energy could be substantially higher. For example, with a true neutrino energy of $\sim$1 PeV, we would instead require photons $E_{\gamma} > $8 eV for pion production.

During pion production roughly half of the energy will be lost through the neutrino-less $\pi^{0}$ channel\cite{2010ApJ...721..630H}, while for the charged pion channel energy is shared roughly equally among the decay products $\pi^{\pm} \rightarrow e^{\pm} + \overset{\brobor}{\nu_{e}} + \overset{-}{\nu_{\mu}} + \nu_{\mu}$\cite{Waxman:1998yy}. Thus $\sim$3/8 of the pion energy is transferred to neutrinos, with a 1:2:0 flavour composition at source. However, across the cosmological baseline travelled, neutrino oscillations will lead to a mixed 1:1:1 composition on Earth. The IceCube realtime event selection is dominated by muon neutrinos, a channel which will carry no more than $\sim$1/8 of the pionic energy. Thus we find:

\begin{equation}
E_{\nu} \approx f_{\pi} \frac{E_{p}}{8}
\end{equation} where $f_{\pi} \leq1$ is the conversion efficiency of proton energy to pion energy. We can derive the mean free path, $\lambda$, for a proton:

\begin{equation}
\lambda = \frac{1}{\sigma_{p\gamma} n_{\gamma}}
\end{equation} with cross section $\sigma_{p\gamma} \sim 5 \times 10^{-28}$ cm$^{2}$ and photon number density $n_{\gamma}$. For a blackbody of temperature $T_{BB} \sim 10^{4.6}$ K, the mean free path for the parent proton of a 1 PeV neutrino is $\lambda \sim 2 \times 10^{13}$ cm. Accounting for the fact that each proton interaction will lead to a typical energy reduction of 20\%, we then find:

\begin{equation}
f_{\pi}(x) = 1 - e^{\left( \frac{-0.2x}{\lambda} \right)}
\end{equation} for path $x$. Equating $x$ with the radius of the UV photosphere $x \approx 10^{14.6}$cm, we then find that each proton or neutron will typically undergo $\sim$ 10 interactions, which would represent a high efficiency $f_{\pi} \sim 0.9$. We caution that this estimate is only approximate, and that detailed numerical simulations are required to accurately calculate the pion production efficiency\cite{2010ApJ...721..630H}.

We then calculate the effective area for a single high-energy neutrino, under the assumption of a mono-energetic neutrino spectrum which approximates the expectation for p$\gamma$ production. The effective area for IceCube varies from 50-200 m$^{2}$ for a 0.2-10 PeV neutrino energy. Below 1 PeV, this corresponds to a roughly-constant threshold of $6 \times 10^{-4}$ erg cm$^{-2}$ for an expectation of one neutrino alert. Given the redshift of AT2019dsg, we find a required total energy in neutrinos $E_{\nu} \approx 4 \times 10^{51}$ erg to produce a single neutrino alert. 
%With the assumed baryon loading ratio and beaming fraction, 
We can thus express the expected number of detected neutrinos as:

\begin{equation}
N_\nu \approx  0.03 \times \frac{E_{\nu}}{10^{50} \rm erg}
\end{equation} 

This expectation would also be valid for any power-law distribution in the same energy range.

\section{Compatibility with existing constraints}
\label{sec:diffuse}

We can estimate the contribution of TDEs to the diffuse neutrino flux that would be required to produce an observation of one association with our ZTF follow-up program. As outlined in Table \ref{tab:nu_alerts}, a total of eight neutrino alerts were observed. For all but one of these, IceCube reported an estimate of the \emph{signalness}, i.e., the probability for each to be astrophysical. We note that this quantity is not an absolute value, but is rather derived under specific assumptions about the underlying neutrino source population. Nonetheless, if we take these estimates at face value, and assume that the additional event had the reported signalness mean of 0.5, we would expect a total of $\sim$4.3 astrophysical neutrinos in our sample. Taking the implied ZTF population expectation of $0.05 < N_{\nu, \textup{tot}} < 4.74$, we would then require that a fraction $0.01 < f < 1.00$ of the astrophysical neutrino flux was produced by ZTF-detected TDEs at 90\% confidence.

\begin{figure}
    \centering
    \includegraphics[width=\textwidth]{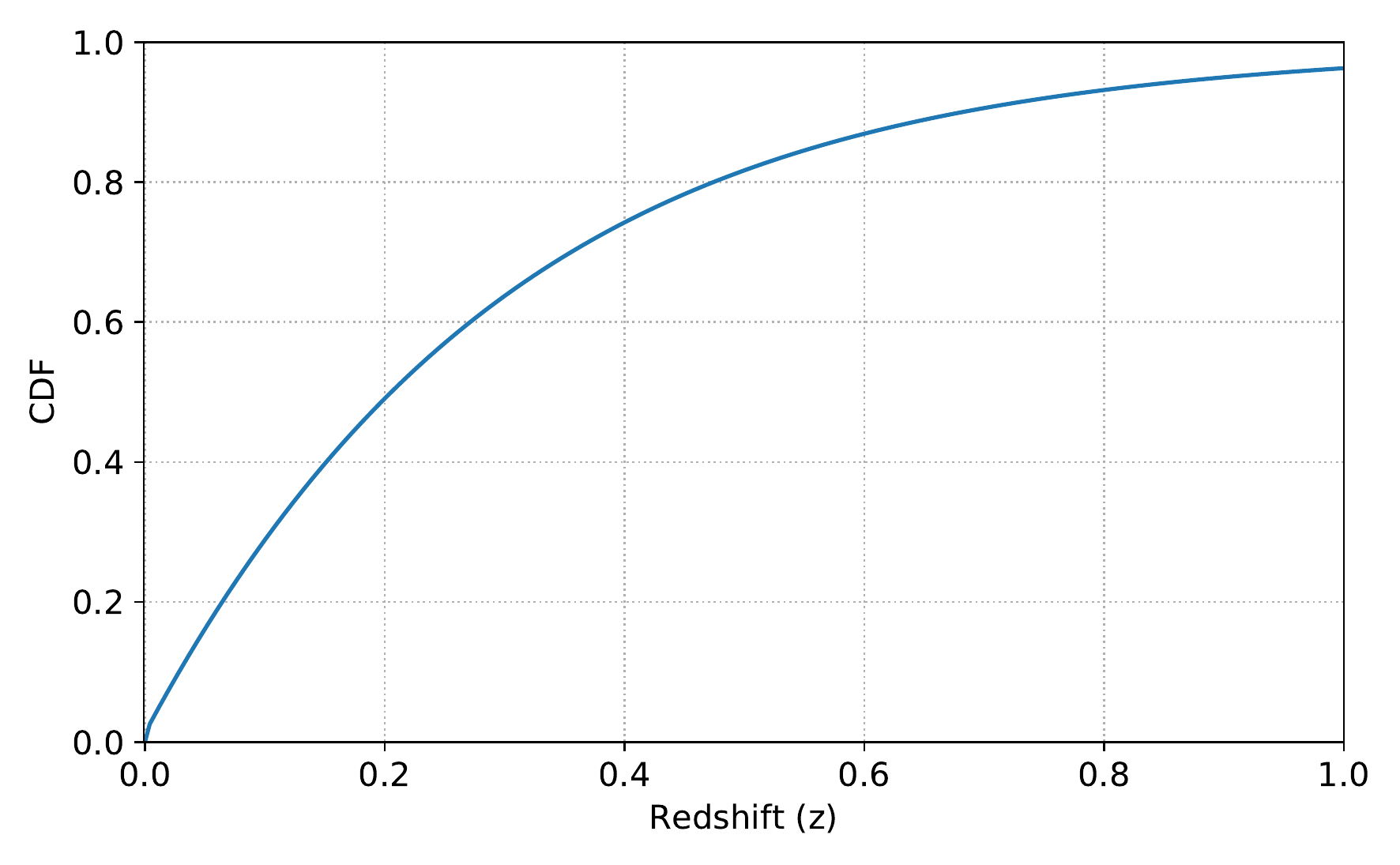}
    \caption{\textbf{Cumulative distribution function (CDF) for TDE neutrino emission as a function of redshift.} These values are derived under the assumption of an E$^{-2.5}$ power law and a negative source evolution\cite{Sun:2015bda}. Roughly half of all neutrinos will come from sources with $z<0.2$.
    }
    \label{fig:cdf}
\end{figure}

We can further consider the contribution of those TDEs that ZTF has not detected, to  estimate the cumulative contribution of all TDEs to the diffuse neutrino flux. The IceCube collaboration has already constrained the contribution of such TDEs to be less than 39\% of the total, under the assumption of an unbroken E$^{-2.5}$ power law and a negative source evolution \cite{2019ICRC...36.1016S, Sun:2015bda}. We follow the same convention, with any power-law contribution of a transient source population to the diffuse neutrino flux is given by:

\begin{equation}
\frac{dN(E, z_{max})}{dEdAdt} = \int_{0}^{z_{max}} \left[ \ (1+z)^{2 - \gamma} \times \frac{\rho(z)\phi_{0}}{4 \pi D_{L}^{2}} \times \left( \frac{E}{E_{0}}\right) ^{-\gamma}  \right] \frac{dV_{C}}{dz} dz
\label{eq:nu_flux_tot}
\end{equation}

where $\rho(z)$ is the source rate density, $\phi_{0}$ is the time-integrated particle flux normalisation at reference energy $E_{0}$. We can use this to calculate the cumulative distribution of neutrino flux as a function of redshift\cite{flarestack}. This CDF is illustrated in Figure \ref{fig:cdf} for an E$^{-2.5}$ power law, though the distribution only depends weakly on the assumed neutrino spectrum.

It is clear that, for this negative source evolution, the vast majority of TDE neutrinos are expected to arrive from sources in the local universe. This statement is independent of both the overall level of TDE neutrino production and the absolute TDE rate. ZTF has, thus far, reported the detection of TDEs up to a maximum redshift 0.212\cite{2020arXiv200101409V}. If we simply assume that ZTF can routinely detect TDEs up to a redshift of 0.15, fully 40\% of the total population flux should come from ZTF-detected sources. We would thus require that at least 2.8\% of the astrophysical neutrino alerts are produced by TDEs, which is fully compatible with the previous IceCube limit. TDE-neutrino associations can thus be detected even if the vast majority of the astrophysical neutrino flux is produced by other source classes. The large relative contribution of detectable TDEs to the population neutrino flux is in marked contrast to supernova-like populations which are dominated by distant sources\cite{ps_iceube_19}, so follow-up searches for TDEs are significantly more sensitive than for these other potential sources.

In any case, IceCube limits are derived under the assumptions of unbroken power laws which extend across a broad energy range (100 GeV - 10 PeV), where many additional neutrinos would be expected at lower energies.  However, for the case of neutrino spectra dominated by high-energy components (as expected for p$\gamma$ neutrino production), no such low-energy neutrinos would be expected, and these existing constraints would then be substantially weakened.

{\bf Supplementary References}\\[-5pt]
\bibliography{supplementary_information.bib}
\newpage

\begin{table*}
 \centering
 	\begin{tabular}{||c c c c c c c ||} 
 		\hline
 		\textbf{Event} & \textbf{R.A. (J2000)} & \textbf{Dec (J2000)} & \textbf{90\% area} & \textbf{ZTF obs} &~ \textbf{Signalness}& \textbf{Ref}\\
 		& \textbf{(deg)}&\textbf{(deg)}& \textbf{(sq. deg.)}& \textbf{(sq. deg.)} &&\\
 		\hline
 		IC190503A & 120.28 & +6.35 & 1.94& 1.37 & 36\%&\cite{blaufuss:gcn24378,2019ATel12730....1S}\\
 		IC190619A & 343.26 & +10.73 & 27.16& 21.57 & 55\%&\cite{blaufuss:gcn24910, 2019ATel12879....1S}\\
 		IC190730A & 225.79 & +10.47 & 5.41& 4.52 & 67\%&\cite{stein:gcn25225,2019ATel12974....1S}\\
 		IC190922B & 5.76 & -1.57 & 4.48 & 4.09 & 51\%&\cite{blaufuss:gcn25806,2019ATel13125....1S, stein:gcn25824}\\
 		\textbf{IC191001A} & \textbf{314.08} & \textbf{+12.94} & \textbf{25.53} & \textbf{20.56} & \textbf{59\%}& \textbf{\cite{stein:gcn25913,2019ATel13160....1S, stein:gcn25929}}\\
 		IC200107A & 148.18 & +35.46 & 7.62 & 6.22 & - &\cite{stein:gcn26655,stein:gcn26667}\\
 		IC200109A & 164.49 & +11.87 & 22.52 & 20.06 & 77\%&\cite{stein:gcn26696,reusch:gcn26747}\\
 		IC200117A & 116.24 & +29.14 & 2.86 &  2.66 & 38\%&\cite{lagunas:gcn26802,reusch:gcn26813, reusch:gcn26816}\\
 		\hline
 	\end{tabular}
 	\caption{\textbf{Summary of the eight neutrino alerts followed up by ZTF.} IC191001A is highlighted in bold. The 90\% area column indicates the region of sky observed at least twice by ZTF, within the reported 90\% localisation, and accounting for chip gaps. The \textit{signalness} estimates the probability that each neutrino is of astrophysical origin, rather than arising from atmospheric backgrounds. One alert, IC200107A, was reported without a signalness estimate.}
 	\label{tab:nu_alerts}
 \end{table*}
 
 \begin{table*}
 \centering
 	\begin{tabular}{||c c ||} 
 		\hline
 		\textbf{Cause} & \textbf{Events} \\
 		\hline
 		\textbf{Alert Retraction} & IC180423A\cite{IC180423A}, IC181031A\cite{IC181031A}, IC190205A\cite{IC190205A}, IC190529A\cite{IC190529A}\\
 		\hline
 		\textbf{Proximity to Sun} &IC180908A\cite{IC180908A}, IC181014A\cite{IC181014A}, IC190124A\cite{IC190124A}, IC190704A\cite{IC190704A}\\
 		& IC190712A\cite{IC190712A}, IC190819A\cite{IC190819A}, IC191119A\cite{IC191119A}, IC200227A\cite{IC200227A}\\
 		\textbf{Low Altitude} & IC191215A\cite{IC191215A}\\
 		\textbf{Southern Sky} & IC190331A\cite{IC190331A}, IC190504A\cite{IC190504A}\\
 		\hline
 		\textbf{Poor Signalness \& Localisation} &
 		IC190221A\cite{IC190221A}, IC190629A\cite{IC190629A}, IC190922A\cite{IC190922A}\\
 		& IC191122A\cite{IC191122A}, IC191204A\cite{IC191204A}, IC191231A\cite{IC191231A}\\
 		\hline
 		\textbf{Bad Weather} & IC200120A\cite{IC200120A,IC200120A_2}\\
 		\textbf{Telescope Maintenance} & IC181023A\cite{IC181023A}\\
 		\hline
 	\end{tabular}
 	\caption{\textbf{Summary of the 23 neutrino alerts that were not followed up by ZTF since survey start on 2018 March 20.} Of these, 4/23 were retracted, 11/23 were inaccessible to ZTF for various reasons, 6/23 were deemed alerts of poor quality, while just 2/23 were alerts that were missed although they passed our criteria.}
 \label{tab:nu_non_observed}
\end{table*}

\newpage
\clearpage

\begin{longtable}{|c|c|c|c|c|c|}
\caption{\textbf{Photometry for AT2019dsg, measured by \textit{Swift}-UVOT, ZTF, LT (IOO) and SEDM.} The time ($\Delta t$) is measured in the observer frame relative to MJD 58582.8, the date of discovery for AT2019dsg. \label{tab:photometry}}\\
\hline
\textbf{$\Delta{t}$}&\textbf{Band}&\textbf{Flux [mJy]}&\textbf{Flux Error [mJy]}&\textbf{$\nu$ [$10^{14}$ Hz]}&\textbf{Instrument}\\%
\hline%
\endfirsthead
\hline%
\textbf{$\Delta{t}$}&\textbf{Band}&\textbf{Flux [mJy]}&\textbf{Flux Error [mJy]}&\textbf{$\nu$ [$10^{14}$ Hz]}&\textbf{Instrument}\\%
\hline%
\endhead
\hline
\endfoot
\hline
\endlastfoot
\hline%
34.02&i&0.19&0.02&4.23&LT (IOO)\\%
38.76&i&0.19&0.02&4.23&LT (IOO)\\%
47.32&i&0.17&0.02&4.23&LT (IOO)\\%
67.26&i&0.13&0.01&4.23&LT (IOO)\\%
0.00&r&0.09&0.01&4.96&ZTF\\%
10.45&r&0.15&0.01&4.96&ZTF\\%
18.10&r&0.19&0.02&4.96&ZTF\\%
23.80&r&0.20&0.02&4.96&ZTF\\%
43.70&r&0.19&0.02&4.96&ZTF\\%
49.43&r&0.16&0.01&4.96&ZTF\\%
52.28&r&0.17&0.02&4.96&ZTF\\%
55.17&r&0.15&0.01&4.96&ZTF\\%
58.98&r&0.15&0.01&4.96&ZTF\\%
64.65&r&0.14&0.01&4.96&ZTF\\%
67.50&r&0.13&0.01&4.96&ZTF\\%
71.30&r&0.11&0.01&4.96&ZTF\\%
72.23&r&0.12&0.01&4.96&ZTF\\%
75.03&r&0.11&0.01&4.96&ZTF\\%
76.11&r&0.11&0.01&4.96&ZTF\\%
77.02&r&0.12&0.01&4.96&ZTF\\%
78.92&r&0.11&0.01&4.96&ZTF\\%
81.71&r&0.11&0.01&4.96&ZTF\\%
93.09&r&0.10&0.01&4.96&ZTF\\%
97.01&r&0.10&0.01&4.96&ZTF\\%
103.60&r&0.09&0.01&4.96&ZTF\\%
104.62&r&0.09&0.01&4.96&ZTF\\%
106.44&r&0.08&0.01&4.96&ZTF\\%
109.33&r&0.07&0.01&4.96&ZTF\\%
115.05&r&0.07&0.01&4.96&ZTF\\%
129.25&r&0.05&0.01&4.96&ZTF\\%
163.38&r&0.05&0.01&4.96&ZTF\\%
166.30&r&0.05&0.01&4.96&ZTF\\%
167.13&r&0.05&0.00&4.96&ZTF\\%
32.37&r&0.22&0.02&5.14&SEDM\\%
34.01&r&0.19&0.02&5.14&LT (IOO)\\%
38.76&r&0.17&0.02&5.14&LT (IOO)\\%
47.32&r&0.16&0.01&5.14&LT (IOO)\\%
23.79&g&0.20&0.02&6.67&ZTF\\%
33.29&g&0.19&0.02&6.67&ZTF\\%
43.76&g&0.18&0.02&6.67&ZTF\\%
49.46&g&0.16&0.01&6.67&ZTF\\%
49.48&g&0.17&0.02&6.67&ZTF\\%
52.32&g&0.15&0.01&6.67&ZTF\\%
55.16&g&0.15&0.01&6.67&ZTF\\%
61.83&g&0.15&0.01&6.67&ZTF\\%
64.68&g&0.12&0.01&6.67&ZTF\\%
67.48&g&0.11&0.01&6.67&ZTF\\%
76.06&g&0.10&0.01&6.67&ZTF\\%
76.09&g&0.10&0.01&6.67&ZTF\\%
78.95&g&0.10&0.01&6.67&ZTF\\%
81.79&g&0.09&0.01&6.67&ZTF\\%
87.48&g&0.08&0.01&6.67&ZTF\\%
93.21&g&0.08&0.01&6.67&ZTF\\%
100.70&g&0.08&0.01&6.67&ZTF\\%
103.56&g&0.08&0.01&6.67&ZTF\\%
104.59&g&0.07&0.01&6.67&ZTF\\%
104.59&g&0.07&0.01&6.67&ZTF\\%
104.60&g&0.07&0.01&6.67&ZTF\\%
104.64&g&0.07&0.01&6.67&ZTF\\%
106.42&g&0.07&0.01&6.67&ZTF\\%
120.79&g&0.05&0.01&6.67&ZTF\\%
123.42&g&0.06&0.01&6.67&ZTF\\%
135.90&g&0.04&0.00&6.67&ZTF\\%
142.54&g&0.04&0.01&6.67&ZTF\\%
156.79&g&0.03&0.01&6.67&ZTF\\%
159.57&g&0.03&0.01&6.67&ZTF\\%
163.42&g&0.03&0.00&6.67&ZTF\\%
166.20&g&0.03&0.00&6.67&ZTF\\%
166.22&g&0.03&0.00&6.67&ZTF\\%
167.16&g&0.03&0.00&6.67&ZTF\\%
168.12&g&0.03&0.00&6.67&ZTF\\%
34.01&g&0.19&0.02&6.8&LT (IOO)\\%
38.76&g&0.19&0.02&6.8&LT (IOO)\\%
47.32&g&0.17&0.02&6.8&LT (IOO)\\%
67.26&g&0.12&0.01&6.8&LT (IOO)\\%
71.07&g&0.11&0.01&6.8&LT (IOO)\\%
74.85&g&0.11&0.01&6.8&LT (IOO)\\%
35.90&B&0.34&0.04&7.31&Swift{-}UVOT\\%
39.59&B&0.24&0.06&7.31&Swift{-}UVOT\\%
42.56&B&0.23&0.06&7.31&Swift{-}UVOT\\%
45.53&B&0.22&0.06&7.31&Swift{-}UVOT\\%
48.60&B&0.19&0.05&7.31&Swift{-}UVOT\\%
51.21&B&0.21&0.05&7.31&Swift{-}UVOT\\%
54.59&B&0.17&0.04&7.31&Swift{-}UVOT\\%
67.65&B&0.26&0.06&7.31&Swift{-}UVOT\\%
35.90&U&0.31&0.02&9.18&Swift{-}UVOT\\%
39.59&U&0.27&0.04&9.18&Swift{-}UVOT\\%
42.56&U&0.29&0.04&9.18&Swift{-}UVOT\\%
45.53&U&0.31&0.03&9.18&Swift{-}UVOT\\%
48.60&U&0.27&0.03&9.18&Swift{-}UVOT\\%
48.69&U&0.23&0.03&9.18&Swift{-}UVOT\\%
51.02&U&0.23&0.03&9.18&Swift{-}UVOT\\%
51.21&U&0.24&0.03&9.18&Swift{-}UVOT\\%
54.59&U&0.23&0.02&9.18&Swift{-}UVOT\\%
57.05&U&0.23&0.02&9.18&Swift{-}UVOT\\%
59.89&U&0.18&0.02&9.18&Swift{-}UVOT\\%
63.10&U&0.24&0.03&9.18&Swift{-}UVOT\\%
67.65&U&0.18&0.03&9.18&Swift{-}UVOT\\%
67.99&U&0.19&0.02&9.18&Swift{-}UVOT\\%
88.18&U&0.15&0.03&9.18&Swift{-}UVOT\\%
88.93&U&0.11&0.02&9.18&Swift{-}UVOT\\%
93.09&U&0.11&0.02&9.18&Swift{-}UVOT\\%
175.15&U&0.06&0.01&9.18&Swift{-}UVOT\\%
179.76&U&0.05&0.02&9.18&Swift{-}UVOT\\%
184.37&U&0.04&0.02&9.18&Swift{-}UVOT\\%
35.90&UVW1&0.38&0.02&12.6&Swift{-}UVOT\\%
39.59&UVW1&0.36&0.03&12.6&Swift{-}UVOT\\%
42.56&UVW1&0.34&0.02&12.6&Swift{-}UVOT\\%
45.52&UVW1&0.33&0.02&12.6&Swift{-}UVOT\\%
48.60&UVW1&0.34&0.02&12.6&Swift{-}UVOT\\%
48.69&UVW1&0.29&0.02&12.6&Swift{-}UVOT\\%
51.02&UVW1&0.28&0.02&12.6&Swift{-}UVOT\\%
51.21&UVW1&0.30&0.02&12.6&Swift{-}UVOT\\%
54.59&UVW1&0.30&0.02&12.6&Swift{-}UVOT\\%
57.05&UVW1&0.27&0.01&12.6&Swift{-}UVOT\\%
59.89&UVW1&0.26&0.02&12.6&Swift{-}UVOT\\%
63.10&UVW1&0.28&0.02&12.6&Swift{-}UVOT\\%
67.65&UVW1&0.26&0.02&12.6&Swift{-}UVOT\\%
67.99&UVW1&0.20&0.01&12.6&Swift{-}UVOT\\%
88.17&UVW1&0.17&0.02&12.6&Swift{-}UVOT\\%
88.93&UVW1&0.15&0.02&12.6&Swift{-}UVOT\\%
93.09&UVW1&0.15&0.01&12.6&Swift{-}UVOT\\%
105.73&UVW1&0.13&0.01&12.6&Swift{-}UVOT\\%
107.43&UVW1&0.11&0.01&12.6&Swift{-}UVOT\\%
111.17&UVW1&0.10&0.01&12.6&Swift{-}UVOT\\%
115.15&UVW1&0.11&0.01&12.6&Swift{-}UVOT\\%
118.88&UVW1&0.12&0.01&12.6&Swift{-}UVOT\\%
123.99&UVW1&0.09&0.01&12.6&Swift{-}UVOT\\%
169.90&UVW1&0.06&0.01&12.6&Swift{-}UVOT\\%
175.15&UVW1&0.06&0.01&12.6&Swift{-}UVOT\\%
179.76&UVW1&0.06&0.01&12.6&Swift{-}UVOT\\%
184.37&UVW1&0.04&0.01&12.6&Swift{-}UVOT\\%
194.09&UVW1&0.07&0.02&12.6&Swift{-}UVOT\\%
199.71&UVW1&0.07&0.01&12.6&Swift{-}UVOT\\%
223.47&UVW1&0.03&0.01&12.6&Swift{-}UVOT\\%
329.98&UVW1&0.03&0.01&12.6&Swift{-}UVOT\\%
35.91&UVM2&0.37&0.01&14.15&Swift{-}UVOT\\%
39.59&UVM2&0.35&0.02&14.15&Swift{-}UVOT\\%
42.57&UVM2&0.35&0.02&14.15&Swift{-}UVOT\\%
45.53&UVM2&0.35&0.02&14.15&Swift{-}UVOT\\%
48.60&UVM2&0.32&0.01&14.15&Swift{-}UVOT\\%
48.69&UVM2&0.33&0.02&14.15&Swift{-}UVOT\\%
51.03&UVM2&0.26&0.01&14.15&Swift{-}UVOT\\%
51.22&UVM2&0.29&0.02&14.15&Swift{-}UVOT\\%
54.60&UVM2&0.28&0.01&14.15&Swift{-}UVOT\\%
57.06&UVM2&0.28&0.01&14.15&Swift{-}UVOT\\%
59.90&UVM2&0.26&0.01&14.15&Swift{-}UVOT\\%
63.10&UVM2&0.25&0.01&14.15&Swift{-}UVOT\\%
67.65&UVM2&0.26&0.02&14.15&Swift{-}UVOT\\%
68.00&UVM2&0.20&0.01&14.15&Swift{-}UVOT\\%
88.18&UVM2&0.15&0.01&14.15&Swift{-}UVOT\\%
88.94&UVM2&0.18&0.01&14.15&Swift{-}UVOT\\%
93.10&UVM2&0.13&0.01&14.15&Swift{-}UVOT\\%
105.72&UVM2&0.12&0.01&14.15&Swift{-}UVOT\\%
107.43&UVM2&0.10&0.01&14.15&Swift{-}UVOT\\%
111.16&UVM2&0.10&0.01&14.15&Swift{-}UVOT\\%
115.15&UVM2&0.10&0.01&14.15&Swift{-}UVOT\\%
118.87&UVM2&0.09&0.01&14.15&Swift{-}UVOT\\%
123.99&UVM2&0.07&0.01&14.15&Swift{-}UVOT\\%
169.91&UVM2&0.05&0.01&14.15&Swift{-}UVOT\\%
175.16&UVM2&0.04&0.00&14.15&Swift{-}UVOT\\%
179.77&UVM2&0.05&0.01&14.15&Swift{-}UVOT\\%
184.38&UVM2&0.04&0.01&14.15&Swift{-}UVOT\\%
194.09&UVM2&0.05&0.01&14.15&Swift{-}UVOT\\%
199.72&UVM2&0.04&0.01&14.15&Swift{-}UVOT\\%
223.48&UVM2&0.04&0.01&14.15&Swift{-}UVOT\\%
329.97&UVM2&0.01&0.00&14.15&Swift{-}UVOT\\%
35.90&UVW2&0.52&0.01&15.69&Swift{-}UVOT\\%
39.59&UVW2&0.47&0.03&15.69&Swift{-}UVOT\\%
42.56&UVW2&0.43&0.02&15.69&Swift{-}UVOT\\%
45.53&UVW2&0.41&0.02&15.69&Swift{-}UVOT\\%
48.60&UVW2&0.42&0.02&15.69&Swift{-}UVOT\\%
48.69&UVW2&0.44&0.02&15.69&Swift{-}UVOT\\%
51.02&UVW2&0.40&0.02&15.69&Swift{-}UVOT\\%
51.21&UVW2&0.39&0.02&15.69&Swift{-}UVOT\\%
54.59&UVW2&0.38&0.01&15.69&Swift{-}UVOT\\%
57.05&UVW2&0.36&0.01&15.69&Swift{-}UVOT\\%
59.89&UVW2&0.36&0.01&15.69&Swift{-}UVOT\\%
63.10&UVW2&0.32&0.02&15.69&Swift{-}UVOT\\%
67.65&UVW2&0.29&0.02&15.69&Swift{-}UVOT\\%
68.00&UVW2&0.31&0.01&15.69&Swift{-}UVOT\\%
88.18&UVW2&0.23&0.01&15.69&Swift{-}UVOT\\%
88.93&UVW2&0.21&0.01&15.69&Swift{-}UVOT\\%
93.10&UVW2&0.20&0.01&15.69&Swift{-}UVOT\\%
105.72&UVW2&0.17&0.01&15.69&Swift{-}UVOT\\%
107.43&UVW2&0.13&0.01&15.69&Swift{-}UVOT\\%
111.16&UVW2&0.12&0.01&15.69&Swift{-}UVOT\\%
115.15&UVW2&0.15&0.01&15.69&Swift{-}UVOT\\%
118.87&UVW2&0.13&0.01&15.69&Swift{-}UVOT\\%
123.98&UVW2&0.10&0.01&15.69&Swift{-}UVOT\\%
169.90&UVW2&0.06&0.01&15.69&Swift{-}UVOT\\%
175.15&UVW2&0.06&0.00&15.69&Swift{-}UVOT\\%
179.77&UVW2&0.06&0.01&15.69&Swift{-}UVOT\\%
184.38&UVW2&0.07&0.01&15.69&Swift{-}UVOT\\%
194.09&UVW2&0.06&0.01&15.69&Swift{-}UVOT\\%
199.72&UVW2&0.04&0.01&15.69&Swift{-}UVOT\\%
223.47&UVW2&0.06&0.01&15.69&Swift{-}UVOT\\%
329.96&UVW2&0.02&0.00&15.69&Swift{-}UVOT\\%
\end{longtable}

\newpage

\begin{longtable}{|c|c|c|c|c|}
\caption{\textbf{X-ray observations of AT2019dsg from \textit{Swift}-XRT and \textit{XMM-Newton}.} The time ($\Delta t$) is measured in the observer frame relative to MJD 58582.8. After $\Delta t = 65.96$, the source was not detected. For these observations, we instead report $3\sigma$ upper limits.\label{tab:xray_data}}\\
\hline
\textbf{$\Delta t$}&\textbf{Energy Flux} &\textbf{Flux Err}&\textbf{Energy Range}&\textbf{Instrument}\\
& \textbf{[10$^{-12}$ erg cm$^{-2}$ s$^{-1}$]} & \textbf{[10$^{-12}$ erg cm$^{-2}$ s$^{-1}$]} & \textbf{[keV]}& \\
\hline%
\endfirsthead
\hline%
\textbf{$\Delta t$}&\textbf{Energy Flux} &\textbf{Flux Err}&\textbf{Energy Range}&\textbf{Instrument}\\
& \textbf{[10$^{-12}$ erg cm$^{-2}$ s$^{-1}$]} & \textbf{[10$^{-12}$ erg cm$^{-2}$ s$^{-1}$]} & \textbf{[keV]}& \\
\hline%
\endhead
\hline
\endfoot
\hline
\endlastfoot
37.37&4.27&0.42&0.3{-}10&Swift{-}XRT\\%
41.24&1.27&0.67&0.3{-}10&Swift{-}XRT\\%
44.37&1.97&0.46&0.3{-}10&Swift{-}XRT\\%
47.48&3.45&0.61&0.3{-}10&Swift{-}XRT\\%
50.16&1.56&0.04&0.3{-}10&XMM{-}Newton\\%
50.75&2.40&0.34&0.3{-}10&Swift{-}XRT\\%
53.36&1.30&0.26&0.3{-}10&Swift{-}XRT\\%
57.01&0.18&0.10&0.3{-}10&Swift{-}XRT\\%
59.6&0.78&0.23&0.3{-}10&Swift{-}XRT\\%
62.59&0.38&0.15&0.3{-}10&Swift{-}XRT\\%
65.96&0.49&0.23&0.3{-}10&Swift{-}XRT\\%
\hline
70.94&0.37&-&0.3{-}10&Swift{-}XRT\\%
92.72&0.46&-&0.3{-}10&Swift{-}XRT\\%
97.49&0.34&-&0.3{-}10&Swift{-}XRT\\%
110.76&0.78&-&0.3{-}10&Swift{-}XRT\\%
112.56&0.96&-&0.3{-}10&Swift{-}XRT\\%
116.48&0.79&-&0.3{-}10&Swift{-}XRT\\%
120.67&0.64&-&0.3{-}10&Swift{-}XRT\\%
124.59&0.66&-&0.3{-}10&Swift{-}XRT\\%
129.96&0.84&-&0.3{-}10&Swift{-}XRT\\%
146.44&2.99&-&0.3{-}10&Swift{-}XRT\\%
149.09&0.98&-&0.3{-}10&Swift{-}XRT\\%
150.64&0.81&-&0.3{-}10&Swift{-}XRT\\%
178.23&0.66&-&0.3{-}10&Swift{-}XRT\\%
183.68&0.30&-&0.3{-}10&Swift{-}XRT\\%
196.16&0.09&-&0.3{-}10&XMM{-}Newton\\%
\end{longtable}

\newpage

\begin{longtable}{|c|c|c|c|c|}
\caption{Radio observations of AT2019dsg from MeerKAT, VLA, and AMI-LA, grouped into quasi-simultaneous epochs. The time ($\Delta t$) is measured in the observer frame relative to MJD 58582.8, the date of discovery for AT2019dsg. \label{tab:radio_data}}\\
\hline
\textbf{$\Delta t$}&\textbf{$\nu$ [GHz]}&\textbf{Flux density [mJy]}&\textbf{Flux Err [mJy]}&\textbf{Instrument}\\%
\hline%
\endfirsthead
\hline%
\textbf{$\Delta t$}&\textbf{$\nu$ [GHz]}&\textbf{Flux density [mJy]}&\textbf{Flux Err [mJy]}&\textbf{Instrument}\\%
\hline%
\endhead
\hline
\endfoot
\hline
\endlastfoot
    42&   8.49&0.290&0.032&VLA      \\
    42&   9.51&0.408&0.035&VLA      \\
    42&  10.49&0.440&0.036&VLA      \\
    42&  11.51&0.412&0.037&VLA      \\
    41&  15.50&0.464&0.045&AMI      \\
\hline
    70&   1.28&0.104&0.018&MeerKAT  \\
    70&   3.50&0.103&0.028&VLA      \\
    70&   4.49&0.319&0.040&VLA      \\
    70&   5.51&0.324&0.033&VLA      \\
    70&   6.49&0.443&0.037&VLA      \\
    70&   7.51&0.558&0.041&VLA      \\
    70&   8.49&0.680&0.045&VLA      \\
    70&   9.51&0.730&0.047&VLA      \\
    70&  10.49&0.756&0.048&VLA      \\
    70&  11.51&0.771&0.051&VLA      \\
    71&  15.50&0.730&0.076&AMI      \\
\hline
   111&   1.28&0.111&0.017&MeerKAT  \\
   120&   2.24&0.249&0.065&VLA      \\
   120&   2.76&0.345&0.053&VLA      \\
   120&   3.18&0.255&0.049&VLA      \\
   120&   3.69&0.419&0.043&VLA      \\
   120&   4.74&0.698&0.046&VLA      \\
   120&   5.76&0.829&0.053&VLA      \\
   120&   6.69&0.987&0.058&VLA      \\
   120&   7.71&1.117&0.063&VLA      \\
   120&   8.49&1.194&0.067&VLA      \\
   120&   9.51&1.238&0.069&VLA      \\
   120&  10.14&1.310&0.073&VLA      \\
   120&  11.16&1.356&0.075&VLA      \\
   119&  15.50&0.978&0.075&AMI      \\
\hline
   179&   1.28&0.152&0.019&MeerKAT  \\
   178&   2.24&0.351&0.055&VLA      \\
   178&   2.75&0.744&0.054&VLA      \\
   178&   3.24&0.920&0.057&VLA      \\
   178&   3.76&1.032&0.063&VLA      \\
   178&   4.74&1.349&0.073&VLA      \\
   178&   5.76&1.379&0.075&VLA      \\
   178&   6.69&1.285&0.071&VLA      \\
   178&   7.71&1.111&0.062&VLA      \\
   178&   8.49&1.074&0.062&VLA      \\
   178&   9.51&0.921&0.054&VLA      \\
   178&  10.14&0.884&0.053&VLA      \\
   178&  11.16&0.785&0.049&VLA      \\
   179&  15.50&0.676&0.055&AMI      \\
\hline
   235&   1.28&0.178&0.032&MeerKAT  \\
   236&  15.50&0.503&0.047&AMI      \\
% 42&8.49&0.29&0.04&VLA\\%
% 42&9.51&0.41&0.05&VLA\\%
% 42&10.49&0.44&0.05&VLA\\%
% 42&11.51&0.41&0.05&VLA\\%
% 41&15.5&0.46&0.06&AMI{-}LA\\%
% \hline
% 70&1.4&0.1&0.02&MeerKAT\\%
% 70&3.5&0.1&0.03&VLA\\%
% 70&4.49&0.32&0.05&VLA\\%
% 70&5.51&0.32&0.05&VLA\\%
% 70&6.49&0.44&0.05&VLA\\%
% 70&7.51&0.56&0.06&VLA\\%
% 70&8.49&0.68&0.06&VLA\\%
% 70&9.51&0.73&0.07&VLA\\%
% 70&10.49&0.76&0.07&VLA\\%
% 70&11.51&0.77&0.07&VLA\\%
% 71&15.5&0.73&0.1&AMI{-}LA\\%
% \hline
% 111&1.4&0.11&0.02&MeerKAT\\%
% 120&2.24&0.25&0.08&VLA\\%
% 120&2.76&0.34&0.07&VLA\\%
% 120&3.18&0.26&0.06&VLA\\%
% 120&3.69&0.42&0.06&VLA\\%
% 120&4.74&0.7&0.06&VLA\\%
% 120&5.76&0.83&0.07&VLA\\%
% 120&6.69&0.99&0.08&VLA\\%
% 120&7.71&1.12&0.09&VLA\\%
% 120&8.49&1.19&0.09&VLA\\%
% 120&9.51&1.24&0.09&VLA\\%
% 120&10.14&1.31&0.1&VLA\\%
% 120&11.16&1.36&0.1&VLA\\%
% 119&15.5&0.98&0.11&AMI{-}LA\\%
% \hline
% 179&1.4&0.15&0.02&MeerKAT\\%
% 178&2.24&0.35&0.07&VLA\\%
% 178&2.76&0.74&0.08&VLA\\%
% 178&3.24&0.92&0.08&VLA\\%
% 178&3.76&1.03&0.09&VLA\\%
% 178&4.74&1.35&0.1&VLA\\%
% 178&5.76&1.38&0.1&VLA\\%
% 178&6.69&1.28&0.09&VLA\\%
% 178&7.71&1.11&0.08&VLA\\%
% 178&8.49&1.07&0.08&VLA\\%
% 178&9.51&0.92&0.07&VLA\\%
% 178&10.14&0.88&0.07&VLA\\%
% 178&11.16&0.78&0.07&VLA\\%
% 179&15.5&0.68&0.08&AMI{-}LA\\%
% \hline
% 235&1.4&0.18&0.03&MeerKAT\\%
% 236&15.5&0.5&0.06&AMI{-}LA\\%
\end{longtable}